# The role of large-scale magnetic field in the morphology and evolution of extragalactic radio sources.

R.R.Andreasyan

*Abstract*

We discuss a model of formation of extragalactic radio sources when the parent optical galaxy has a large-scale dipolar magnetic field. The study of dynamics of ejected from the central part of optical galaxy clouds of relativistic particles in dipolar magnetic field gives a possibility to explain main morphological features and physical properties of formed extragalactic radio sources. We bring some results of statistical analyses and correlations between physical parameters for more than 500 radio sources. In appendix we present the data of all used extragalactic radio sources with the references for them.

*1. Introduction.* It is well known that the magnetic field has an important role in the dynamics evolution and radiation of extragalactic radio sources. Almost in all known theories of formation and evolution of extragalactic radio sources as a mechanism of radiation in radio waves are suggested synchrotron radiation of relativistic plasma in the magnetic field without of concretization of the large-scale configuration of magnetic field or of the role of field configuration in the observing morphology of extragalactic radio sources. For example in most known Blandford Znajek (1977) mechanism is supposed the existence of large angular momentum and strong magnetic field, parallel to the rotation axes of the Kerr black hole. Relativistic particles are moving along the magnetic field lines and radiating in this field. As there are no magnetic monopoles in the Universe one must consider that the parallel (in the central part of AGN) lines of magnetic field will be closed in any place of galaxy. From the conclusion of symmetry this magnetic field can have a dipolar form. Another interpretation of the morphology of extragalactic radio sources was suggested by Luhmann (1979), what concerns the possibility that the emission arises from the belts of trapped in dipolar magnetic field electrons, encircling the parent galaxy in the same manner as the Van Allen belts encircle the Earth. In (Andreasyan 1984 (hereafter the paper 1)) we have suggested a mechanism of the formation and evolution of extragalactic radio sources in framework of the cosmological conception of V.Ambartsumian (Ambartsumian 1966). This mechanism was as a hybrid of Blandford Znajek and Luhmann mechanisms. But in paper 1 it was done a main suggestion about the magnetic field configuration of host supergiant elliptical galaxy. We conclude that the magnetic field of the host galaxy or AGN has a dipole configuration, with dipole axes parallel to the rotation axes of host galaxy. Extragalactic radio sources are formed from relativistic plasma clouds, ejected from the central part of the optical galaxy and moving in its large-scale, dipole magnetic field. In the frame of suggested mechanism the well-known Fanaroff-Ryley Dichotomy (Fanaroff & Ryley 1974) and many other morphological fetchers finds a very simple physical explanation.

2. *The large scale magnetic field of parent galaxies .* The first main suggestion for the mechanism of the formation and evolution of extragalactic radio sources in paper 1 is the large scale dipolar configuration of magnetic field of parent galaxy, with dipole axes parallel to the rotation axes of host elliptical galaxy. There are many observational evidences that large-scale galactic magnetic fields can have dipolar configuration (for example in NGC4631 (Dumke et al. 1995), or in NGC5775 (Soida et al. 2011), and also in the halo of our Galaxy (Andreasyan & Macarov 1989, Han et al.1997)). The magnetic fields of dipole configuration can be formed and evaluate, for example, in the result of Biermann battery effect (Biermann 1950), in Active Galactic Nucleus (Lesch et all., 1989; Andreasyan 1996). Partly in (Andreasyan 1996) for the evolution of dipolar magnetic field we suggest a model of AGN in agreement with the cosmological conception of V.Ambartsumian. In agreement with this model from the nucleus of Active galaxy there is a permanent ejection of hot plasma, which expands in the fast rotating gaseous medium of the central part of galaxy. Because of the large differences between scattering time of expanding electrons and protons with the rotating medium, in every point of rotating medium the rotation velocity of scattered electrons and protons will correspond to the rotation velocity of their last scattering point and will be different. In the result of forming of circular electric currents in the central part of Active galaxies evaluates dipolar magnetic fields. At the present time there are a lot of observational evidences of existence of the large amount of neutral and ionized gas in the host elliptical galaxies (for example Morganti et al.2003a, Andreasyan et.al.2008), and also of outflows in Radio galaxies (Morganti et al.2003b), which provides the conditions for the working of Biermann battery effect, and evolution of dipolar magnetic fields.

The second suggestion in paper 1 was that the extragalactic radio sources are formed from relativistic plasma clouds, ejected from the central part of the optical galaxy and moving in large-scale dipolar magnetic field of parent galaxy. The behaviors of relativistic plasma cloud, ejected in the direction of the dipole axis, depends on the ratio Q of the kinetic energy density of the plasma to the magnetic field energy density,

i) If the ratio Q is greater than unity ($Q>1$, the kinetic energy density of the plasma is larger than

the magnetic field energy density), the clouds of charged particles, expanding, travels large distances from the optical galaxy, carrying with them the magnetic field lines, as it takes place in many well known models. In this case we expect to observe the more elongated radio images (Fig.1). The directions of the major axes of the radio images will be close to those of rotation axes or the minor axes of the optical galaxies. The magnetic field will be mainly parallel to the radio axes. It can be formed also radio lobes and hot spots near their outer edges, and also well collimated magnetic canals directed from the AGN to the lobs, which can assist for the formation of well collimated jets in case of the secondary permanent ejection of plasma lower energy density. Similar features we observe in extragalactic radio sources of FRII classes (Fanaroff & Riley 1974).

ii) If the ratio Q of energy densities is less than unity (Q<1, the kinetic energy density of the plasma is less than the magnetic field energy density), the charged particles will move along the field lines of the dipole magnetic field of the galaxy. In the beginning of this process (in the younger radio sources) along the dipole axes will be observed the long, jet like features with relatively large opening angles, like to the jets classified as FRI type.

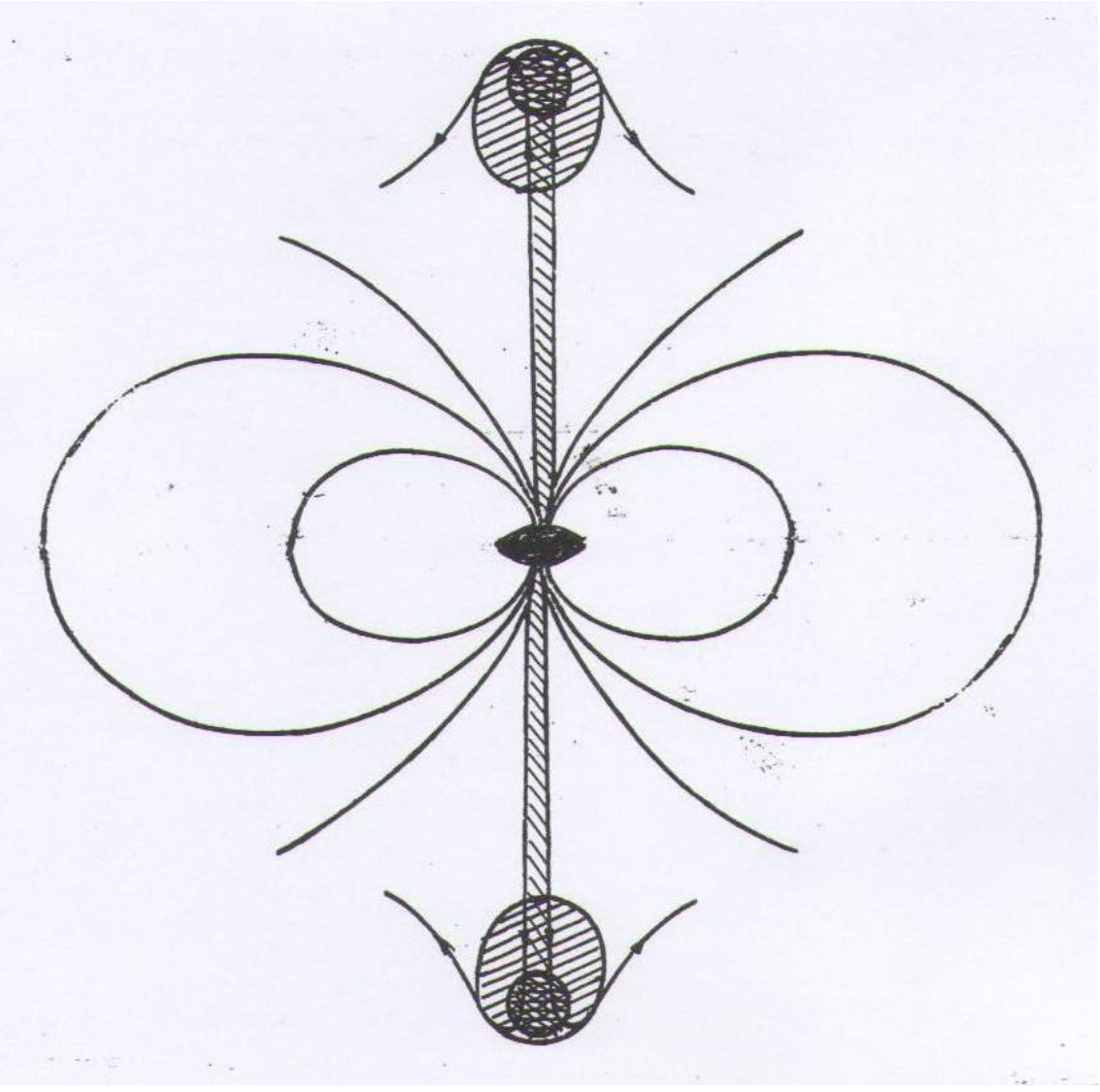

Fig.1. The ratio Q is greater than unity (Q>1).

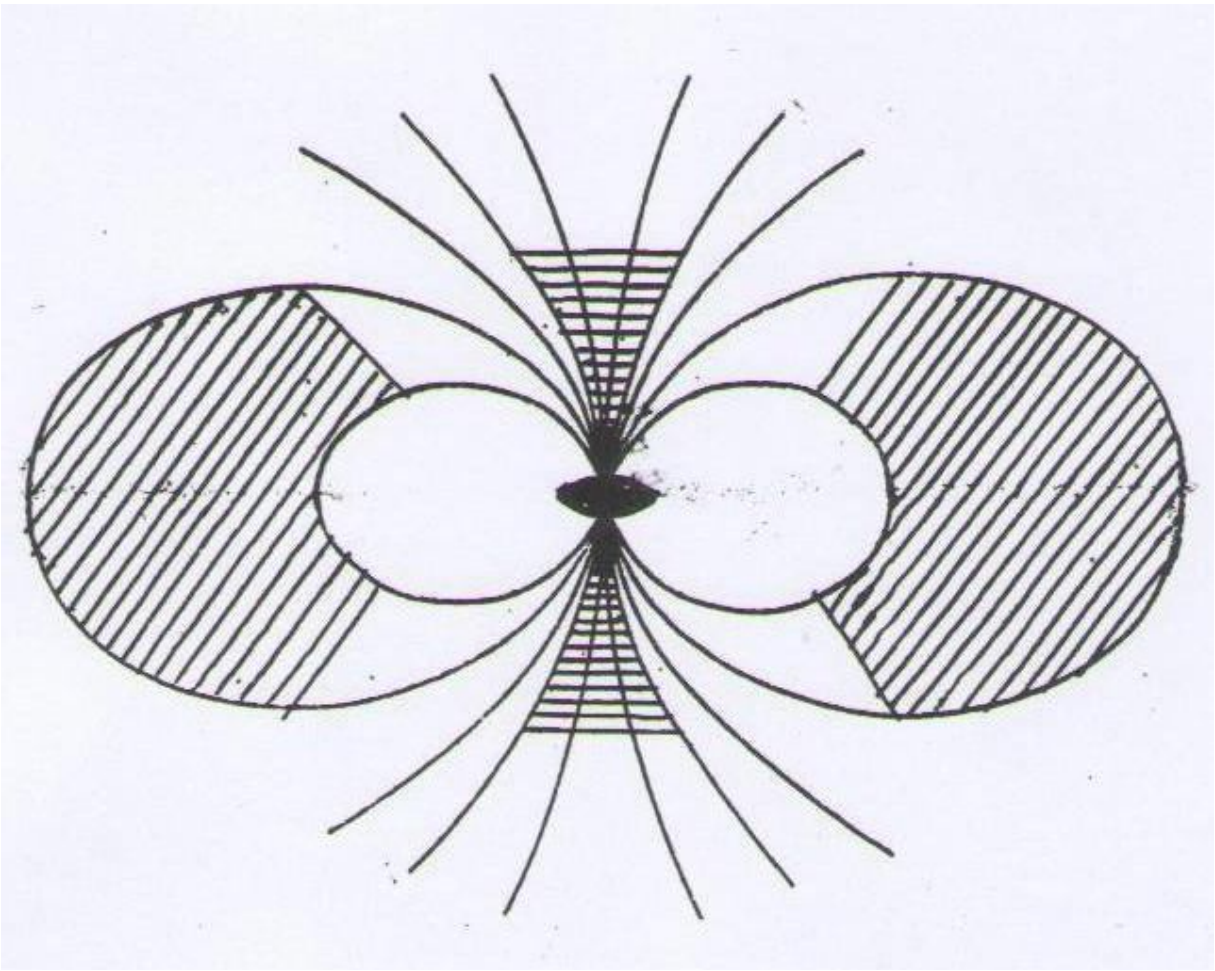

Fig.2. The ratio Q is less than unity (Q<1)

Relativistic charged particles, moving along the field lines of the dipole, after some time will be trapped in the dipolar magnetic field. In result of this all particles finally will execute oscillations in the direction of magnetic lines between magnetic mirrors, and drift in a plane perpendicular to the dipole axis, as it takes place in the Van Allen belts of the Earth (Fig.2). In this case will be formed less elongated radio sources with the radio axes perpendicular to the dipole axes, and correlated with optical major axes. Also we will observe mainly the edge darkened FRI type. As it was shown, in the case of $Q<1$ it can be formed two type edge darkened radio sources; the relatively older Van Allen belts type sources, elongated perpendicular to the dipole axes, and the younger jet like radio sources along the dipole axes. These two types of extragalactic radio sources can be classified as FRI type, though they have different radio orientations relatively to the parent galaxies. It must be noted that such two types of relatively older and younger radio sources with different orientations can be observed near the same optical galaxy. Then we will observe misalignments of radio sources of different size scale, as was found by Appl et al. (1996), or the X shape radio sources (see for example Cheung 2007).

It must be noted that in the literature some authors use the name "Jet" for radio sources that can have different forms and shapes, sometimes for objects that have not any elongation. Here, following to Bridle & Perley (1984), we will use the name "Jet" for the radio source that has the elongation parameter $K>4$, where K is the ratio of the largest dimension of radio image to the perpendicular dimension.

In paper 1 parallel to the well known Fanaroff-Riley classification we bring also a simple classification of extragalactic radio sources by the elongation parameter K. In the case when the charged relativistic particles will be trapped in a dipolar magnetic field, the largest value of the parameter K can be obtained from the equation of the line of force of the dipole field. This ratio is near to 2.5. Thus, one can introduce a quantitative criterion for separating the extragalactic radio sources by their elongation parameter. So, following to this classification, there are:

1) Extragalactic radio sources for which $K>2.5$ in the case of $Q>1$ (FRII type), and for younger jet like radio sources of FRI type in case of $Q<1$;

2) Extragalactic radio sources for which $K<2.5$ in the case of relatively older Van Allen belts type sources ($Q<1$).

As it will be seen in the next paragraphs there are some correlations between FR and this classification by the elongation parameter K. The statistical analyses of observational data were done parallel for the FR and K classification.

*3. The Fanaroff-Riley Dichotomy of extragalactic radio sources.* As it was mentioned above, the Fanaroff-Riley (1974) classification of extragalactic radio sources was done using the morphological features, the edge darkened-FRI, and edge brightened, relatively more luminous FRII types. Probably one can wait also some other morphological and physical differences between the different FR classes of extragalactic radio sources. The study of Fanaroff-Riley (FR) Dichotomy of extragalactic radio sources is very important for understanding and choosing of the mechanism of their formation and evolution. The FR Dichotomy is studying now very intensively, and there are found many other observational differences between the physical properties of these two morphological classes: in the total luminosity, in radio core powers, in ratio of core to lobe radio power, in the relationships between emission-line luminosity and radio power etc (Zirbel & Baum 1995; Gopal-Krishna & Wiita 2000; Gendre 2011, et.cetra). From the mechanism of formation of extragalactic radio sources, discussed above, it is clear that there can be a lot of other differences between different classes of extragalactic radio sources. Here we bring some other observational differences between the different FR types as well as between the types classified by their elongation parameter K.

4. *Observational data.* For this study we have used data for 267 nearby radio galaxies identified with elliptical galaxies brighter than 18$^{th}$ magnitude (sample1) (Andreasyan & Sol, 1999), and 280 extragalactic radio sources with known position angles between the integrated intrinsic radio polarization and radio axes (sample 2) (Andreasyan et all., 2002). For nearby radio sources, we have data: on the position angles of the optical images (oPA) of elliptical galaxies, found mainly from the Palomar maps, the position angles of radio image (rPA), angles between optical and radio axes (dPA), and FR classes taken from the literature. The radio galaxies were also classified as a function of their elongation parameter K using the published radio maps. For the objects of sample 2 also are found FR classes and have been determined elongation parameter K. In samples 1 and 2 we have 289 extragalactic radio sources with known both, FR classification and K parameters. Here we bring some results (physical and morphological differences in different classes of extragalactic radio sources) obtained in our early study (Andreasyan & Sol, 1999; Andreasyan & Sol, 2000, Ap., 43, 413; Andreasyan et all., 2002) as well as some new results. The samples 1 and 2 with the references to the used literature we bring in the Appendix.

5. *The correlation of radio axis with the optical axis in nearby radio galaxies.* Data from sample of 267 nearby radio galaxies were used to study the correlations of radio axes with the optical axes of parent galaxies. Were constructed histograms separately for radio galaxies classified by elongation (Fig.3) and for radio galaxies with FR classes (Fig4). On the figures the difference between the radio and optical position angles (dPA) is laid out along the horizontal axis and the number of radio galaxies along the vertical axis.

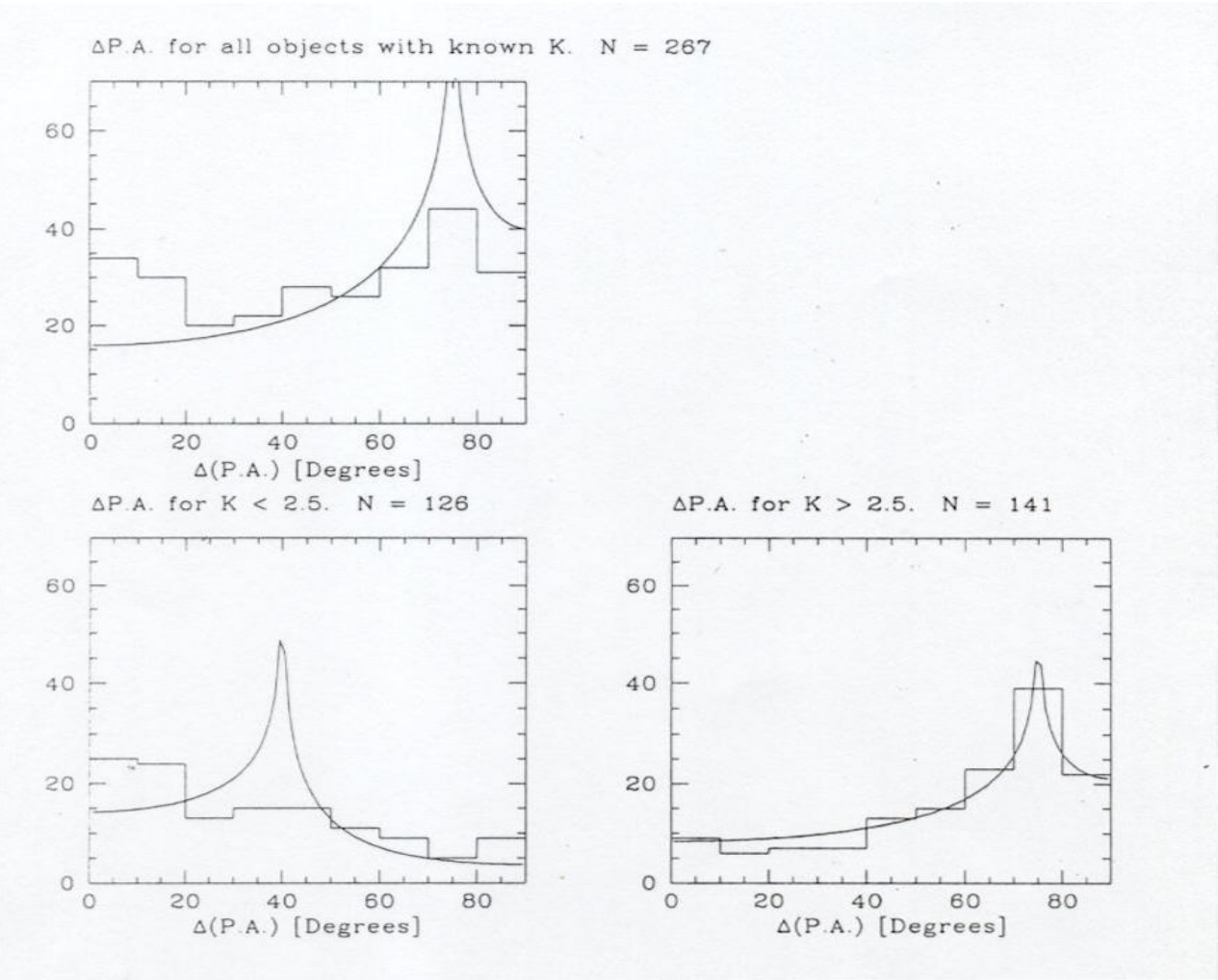

Fig.1. The distribution of (dPA) for K classes.

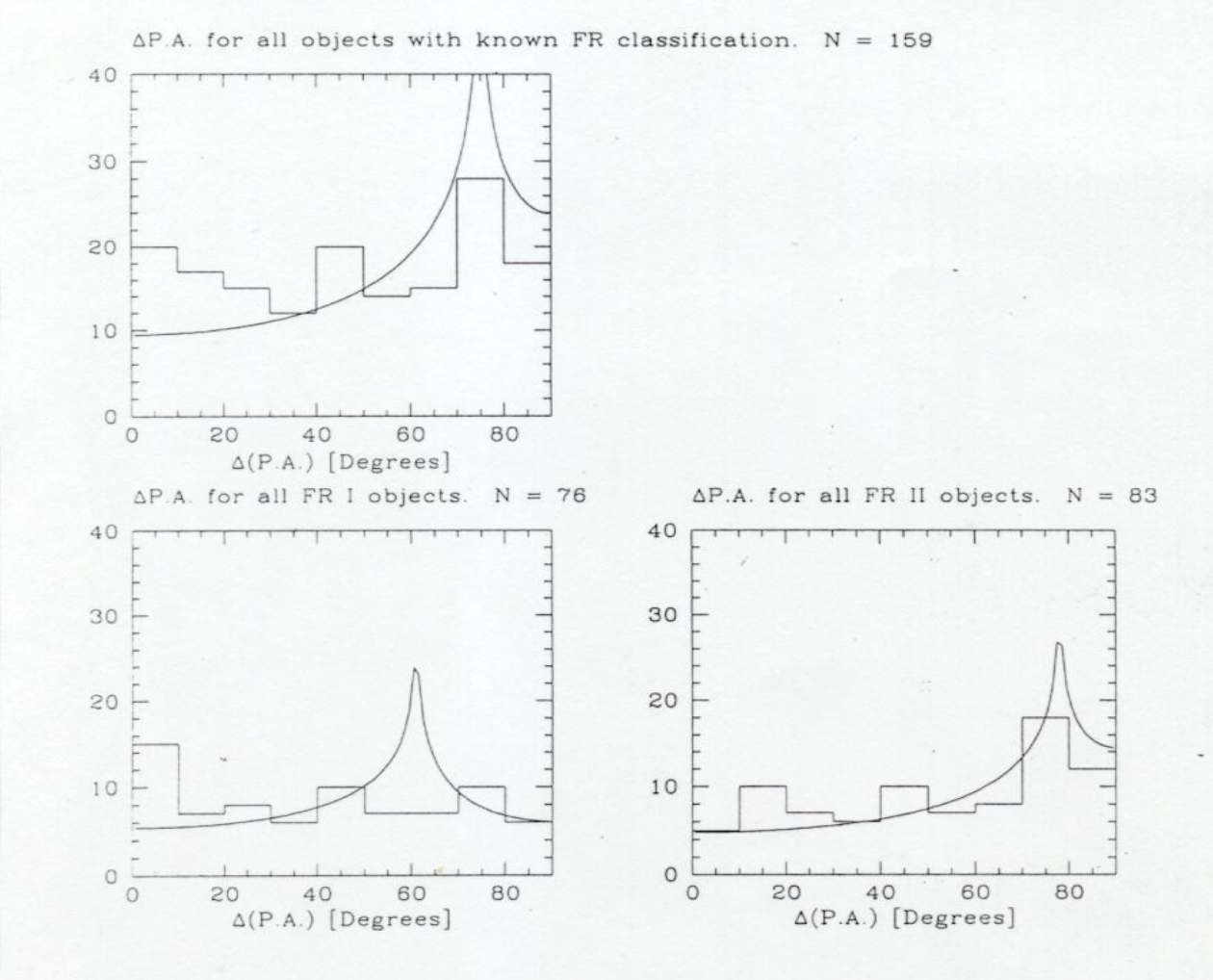

Fig.2. The distribution of (dPA) for FR classes.

On the histograms we bring also the expected distribution of dPA. The continuous lines show the best fits obtained from primary distributions of intrinsic angles, described by a delta-function, taking into account the orientation effect. The fit of observed histograms of these relative orientations have been done using the method developed by Appl et al. (1996).

From the figures we found, that more elongated and FRII type radio galaxies are in most cases directed as minor axes or rotation axes of host elliptical galaxies, while the less elongated and FRI ones are directed perpendicular to these axes. This result is in a good agreement with conclusions of paragraph 2. The weaker correlation for radio sources of FRI type can be explained by our mechanism. As it was shown, if $Q<1$ It can be formatted two type radio sources; the relatively older Van Allen belts type sources (with $K<2.5$), and the younger jet like radio sources along the dipole axes (with $K>2.5$). These two types of extragalactic radio sources are classified as edge darkened FRI type, though they have different radio orientations relatively to parent galaxies

*6. The ellipticity of elliptical galaxies identified with the different types of extragalactic radio sources.* In the Sample1 we have data of the optical ellipticity (E) of 154 elliptical galaxies. For all of them we have the elongation parameters K and for 95 - the Fanaroff-Riley classes. We use this data to study the distribution of ellipticities of parent optical galaxies for different FR types and for different classes of our K classification (Fig.5).

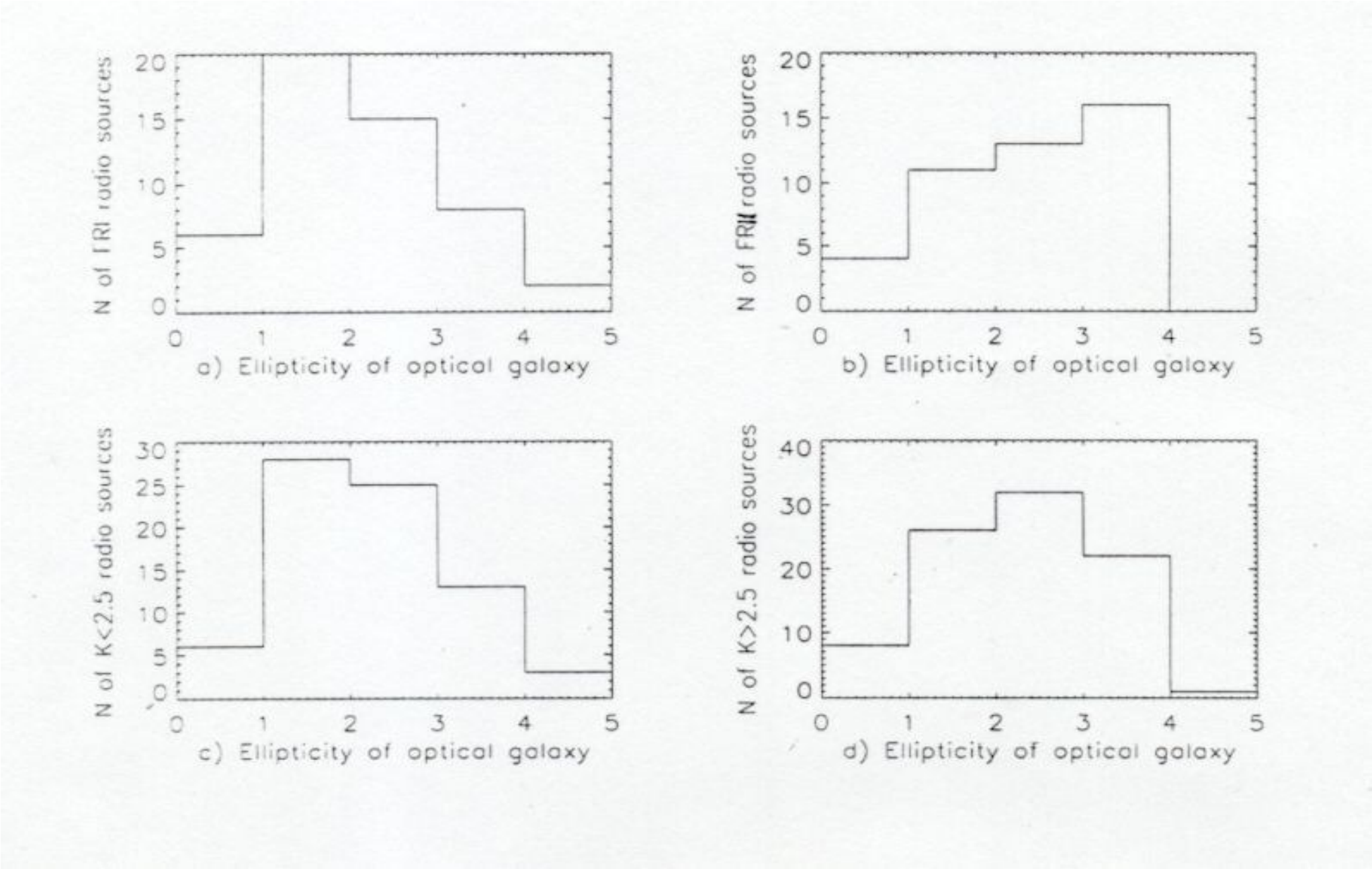

.Fig.5. The ellipticities of optical parent galaxies for different FR types and for different classes of K classification.

From the Fig.5 it is clear that the host elliptical galaxies of less elongated extragalactic radio sources and radio sources of FRI type have less ellipticity (E ~1 to2) than these of radio sources of large elongation and radio galaxies FRII type (E ~3 to 4). The similar result we see for the classification by K parameter. The fact that host elliptical galaxies of FRI type radio sources have less ellipticities can be explained in two ways: It is primordial and in some way is responsible for the formation of FRI types, or it is from the orientation effect. In both cases, the fact of different ellipticities of different FR types is interesting for the understanding of formation of radio sources.

7. *The correlation of the radio polarization angle with the radio axes of extragalactic radio sources.* The data of 280 extragalactic radio sources of sample 2 were used for the study of distribution of angles Δ(PA) between directions of integrated intrinsic radio polarization and the major axes for different type radio sources, classified by their elongation and FR classification. The histograms of angles between radio and polarization axes are shown in (Fig.6) and (Fig.7).

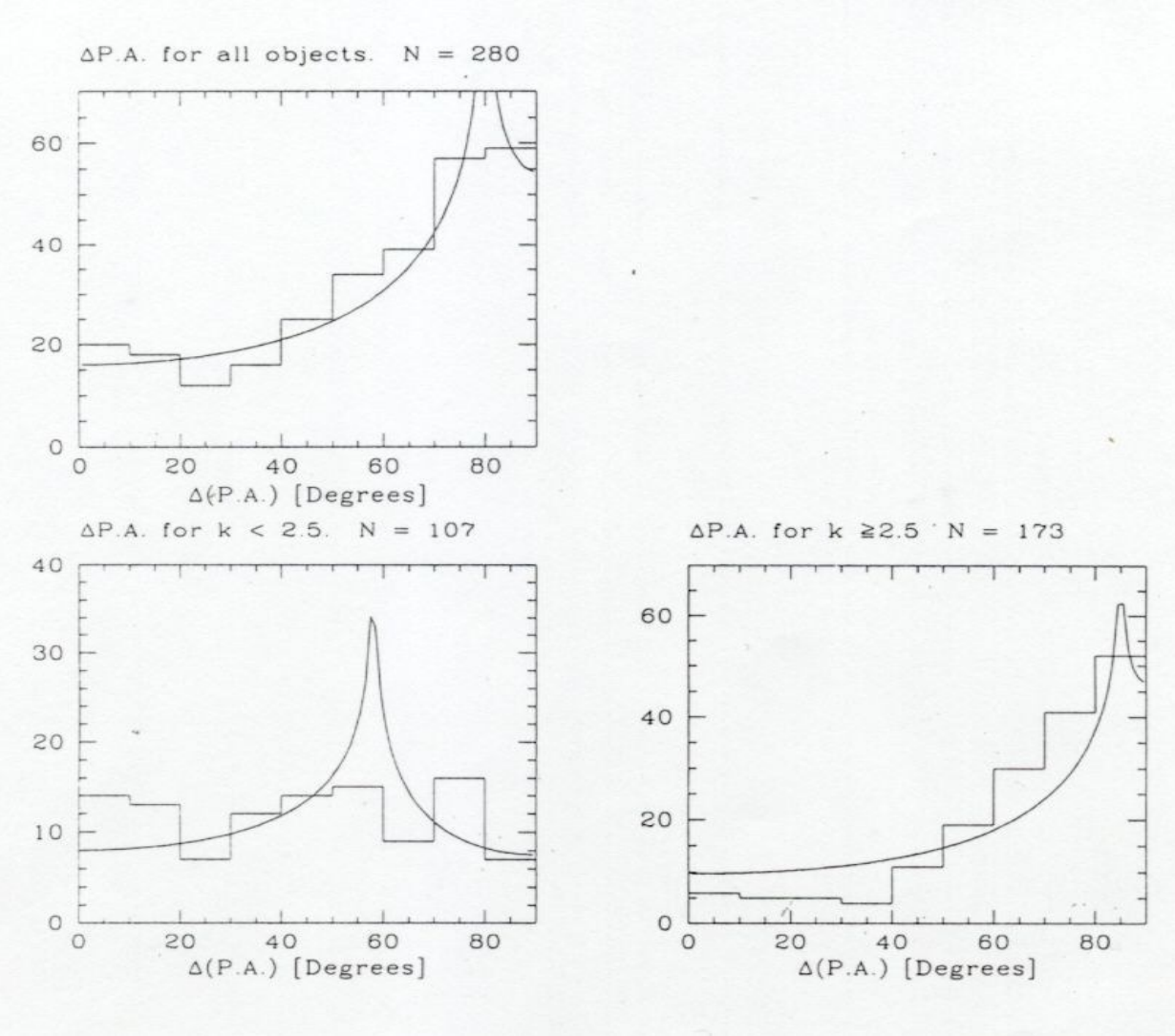

Fig.6. The distribution of Δ(PA) for K classes.

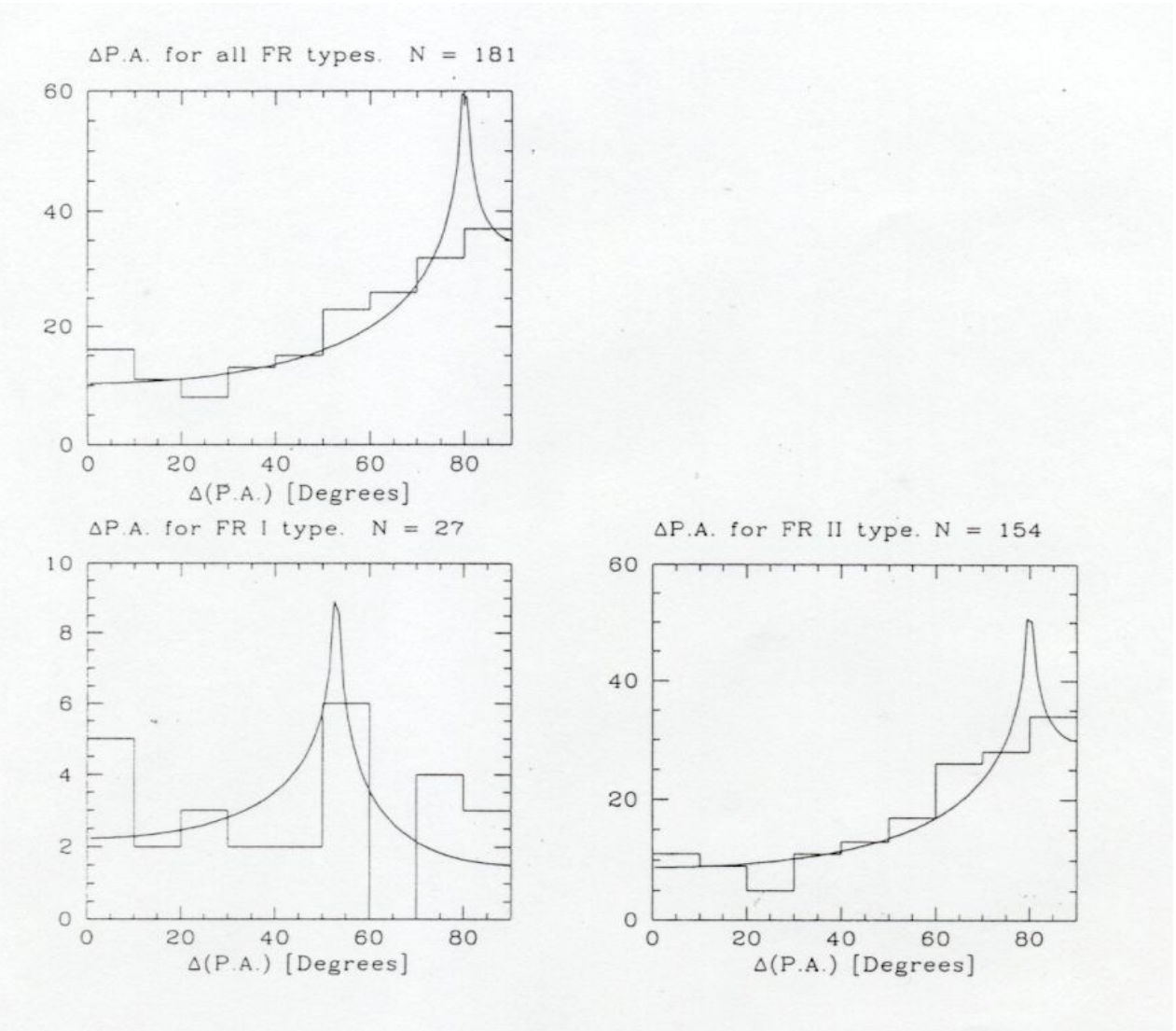

Fig.7. The distribution of Δ(PA) for FR classes.

The fit of histograms for relative orientations also have been done using the method developed by Appl et al. (1996), taking into consideration the projection effects. The continuous line shows the best fits obtained from primary distributions of intrinsic angles described by a delta-function. This method describes rather well the case of elongated and FRII sources, which suggests that their intrinsic integrated polarization is perpendicular to their intrinsic major radio axes. Conversely the less elongated and FRI radio sources do not show any specific intrinsic angle and cannot be fitted by such a simple scenario. As the magnetic fields in optically thin synchrotron radio sources are perpendicular to the polarization of electric vector, the main result of this study is that integrated magnetic fields can be described as intrinsically aligned with major radio axes for elongated and FRII radio sources, while they are not correlated with radio axes for stocky and FRI radio sources.

*8. The relation between FR classes and elongation K parameters.* In samples 1 and 2 we have 289 extragalactic radio sources with known K parameters and FR classes. There are 93 FRI and 196 FRII types of objects. These data was used for the study of the relation of FR classes from the K parameter. In the Fig.8 we bring the distribution function φ(K) of FRI and FRII type radio sources depending from the elongation parameter K.

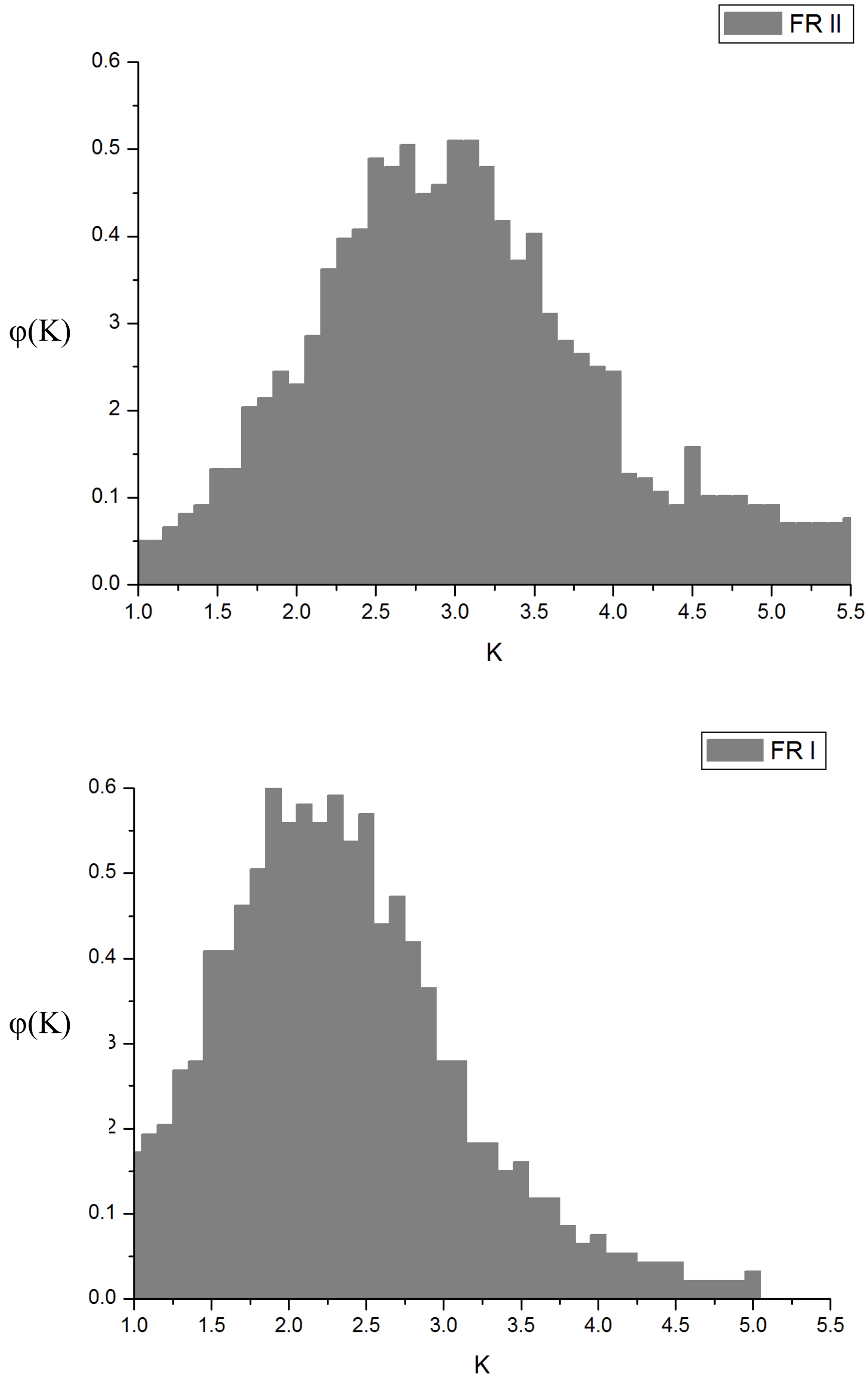

Fig.8. The distribution function of FRII and FRI type radio sources depending from elongation K parameter.

From the fig.8 we see that extragalactic radio sources of FRI and FRII type have different distributions from the K elongation parameter. The maximum of distribution for FRI type is near K=2.25, when it is near K=3.0 for FRII type, and the distribution of FRII type is extended to the large amounts of K. So, extragalactic radio sources of FRII type are more elongated than FRI types.

*9. Conclusions.* The results, obtained above from the analyses of observational data are in good agreement with the suggested mechanism of formation of extragalactic radio sources. Almost all main observed physical and morphological properties of extragalactic radio sources can be qualitatively explained in terms of mentioned scenario, varying the parameter Q as well as the environment of radio sources. It must be noted that mentioned above correlations of FR classification with physical parameters will be better, if we divide the FRI type of extragalactic radio sources in two types: i) the more elongated jet like edge darkened radio sources, and ii) the less elongated (K<2.5, Van Allen belts type sources) and probably older edge darkened radio sources.

Appendix

Tables of extragalactic radio sources: Data for 267 nearby radio galaxies identified with elliptical galaxies brighter than 18$^{th}$ magnitude (sample1), and 280 extragalactic radio sources with known position angles between the integrated intrinsic radio polarization and radio axes (sample 2).

Sample 1.

Col.1: Radio Sours Name

Col.2: oPA is the position angle of the major axis of optical galaxy

Col.3: rPA is the position angle of radio axis

Col.4: dPA is the relative position angle between optical(oPA) and radio(rPA) axis

Col.5: E is the elipticity of the optical galaxy identified with radio sources

Col.6: K (also Col.12) is the ratio of major to minor axis of radio image

Col.7: FR is the Fanaroff Riley classe

Col.8: M is the optical magnitude

Col.9: SI is the spectral index

Col.10: z is the redshift

Col.11: logP is the radio luminocit

Col.13: Ref. for the radio maps and FR classes

| 1 | 2 | 3 | 4 | 5 | 6 | 7 | 8 | 9 | 10 | 11 | 12 | 13 |
|---|---|---|---|---|---|---|---|---|---|---|---|---|
| Object | oPA | rPA | dPA | E | K | FR | M | SI | z | logP (WHz ) | K | ref |
| 0005-199 | 5 | 81 | 76 | 2.8 | >2.5 | | 16.5 | 0.70 | | | 4.2 | 26 |
| 0007+124 | 2 | 21 | 19 | | >2.5 | II | 17.7 | 0.78 | 0.157 | | 2.6 | 20 5 |
| 0013-316 | 39 | 110 | 71 | 1.0 | >2.5 | | 16.5 | 0.92 | | | 3.5 | 26 |
| 0018-194 | 17 | 111 | 86 | 4.0 | <2.5 | II | 17.0 | 0.69 | 0.095 | | 1.4 | 21 |
| 0023-33 | 34 | 70 | 36 | 1.7 | <2.5 | | 16.7 | 0.50 | 0.050 | | 1.6 | 26 |
| 0034+254 | 163 | 83 | 80 | 2.0 | >2.5 | I | 14.8 | 0.66 | 0.032 | 24.07 | 2.8 | 15 15 |
| 0039+211 | 82 | 0 | 82 | 2.0 | >2.5 | | | 0.90 | 0.102 | 24.89 | 2.6 | 56 |
| 0040-06 | 58 | 165 | 73 | | <2.5 | | 17.0 | | | | 1.5 | 21 |
| 0043+201 | 69 | 172 | 77 | 2.0 | >2.5 | | 15.7 | 0.75 | 0.106 | 25.06 | 4.0 | 57 |
| 0043-424 | 159 | 136 | 23 | 2.0 | <2.5 | II | 16.0 | 0.87 | 0.053 | 27.23 | 2.4 | 25 7 |
| 0053+261 | 171 | 146 | 25 | | <2.5 | I | 17.5 | 1.06 | 0.195 | 27.15 | 1.4 | 18 2 |
| 0055+265 | 152 | 109 | 43 | | >2.5 | I | 13.0 | 0.84 | 0.047 | 25.67 | 2.6 | 15 15 |
| 0055+300 | 42 | 129 | 87 | 3.1 | <2.5 | I | 12.2 | 1.04 | 0.017 | 24.52 | 2.4 | 15 15 |
| 0104+321 | 135 | 147 | 12 | | <2.5 | I | 12.1 | 0.57 | 0.017 | 25.07 | 2.0 | 27 2 |
| 0106+130 | 131 | 20 | 69 | 1.0 | >2.5 | II | 15.1 | 0.76 | 0.060 | | 3.4 | 17 2 |
| 0108-142 | 53 | 100 | 47 | 1.0 | >2.5 | I | 15.8 | | 0.052 | 24.96 | 3.2 | 85 7 |
| 0109+492 | 101 | 13 | 88 | 1.0 | >2.5 | II | 15.6 | 0.77 | 0.067 | 26.21 | 4.3 | 18 2 |
| 0110+152 | 105 | 170 | 65 | | >2.5 | | 15.5 | 1.20 | 0.048 | 24.34 | 2.6 | 61 |

| | | | | | | | | | | | | | |
|---|---|---|---|---|---|---|---|---|---|---|---|---|---|
| 0114-476 | 17 | 157 | 40 | | <2.5 | II | 16.5 | 0.60 | 0.146 | 26.86 | 2.2 | 21 | 3 |
| 0116+319 | 50 | 115 | 65 | | <2.5 | | 14.5 | 0.42 | 0.059 | 25.10 | 1.6 | 75 | |
| 0120+33 | 70 | 120 | 50 | | <2.5 | | 13.0 | 1.40 | 0.016 | 23.35 | 2.0 | 15 | |
| 0123-016 | 52 | 178 | 54 | | >2.5 | I | 12.2 | 0.66 | 0.018 | 25.27 | 2.6 | 29 | 15 |
| 0124+189 | 76 | 13 | 63 | 1.0 | >2.5 | | 15.5 | 0.56 | 0.043 | | 2.8 | 85 | |
| 0131-367 | 164 | 89 | 75 | 3.2 | >2.5 | II | 14.2 | 0.51 | 0.030 | 25.76 | 2.6 | 20 | 7 |
| 0149+35 | 30 | 87 | 57 | | >2.5 | I | 14.5 | 0.60 | 0.016 | 23.0 | 5.0 | 15 | 15 |
| 0153+053 | 73 | 84 | 11 | 2.0 | <2.5 | | 13.2 | 0.50 | | | 2.2 | 63 | |
| 0206+355 | 137 | 132 | 5 | | <2.5 | I | 13.0 | 0.66 | 0.037 | 25.44 | 1.5 | 16 | 15 |
| 0214-480 | 100 | 175 | 75 | 1.0 | >2.5 | I | 14.5 | 1.00 | 0.064 | 26.00 | 3.3 | 54 | 7 |
| 0220+427 | 33 | 50 | 17 | | <2.5 | I | 12.5 | 0.50 | 0.022 | 24.69 | 2.0 | 17 | 2 |
| 0222+369 | 80 | 48 | 32 | | <2.5 | | 15.0 | 0.24 | 0.033 | 23.70 | 1.6 | 70 | |
| 0229-208 | 174 | 110 | 64 | 1.6 | >2.5 | II | 16.0 | 0.62 | 0.090 | 25.46 | 4.0 | 26 | |
| 0239-85 | 113 | 95 | 18 | 3.1 | <2.5 | | | | | | 2.0 | 65 | |
| 0247-207 | 45 | 40 | 5 | 0.2 | >2.5 | I | 15.4 | 0.97 | 0.087 | 25.35 | 5.5 | 26 | 7 |
| 0255+133 | 87 | 159 | 72 | 4.0 | >2.5 | II | 16.8 | | 0.075 | 24.07 | 2.6 | 57 | |
| 0257-398 | 115 | 60 | 55 | 2.0 | <2.5 | | 15.3 | 0.64 | | | 1.5 | 26 | |
| 0258+350 | 70 | 126 | 56 | | <2.5 | | 13.5 | 0.54 | 0.016 | 24.63 | 2.2 | 22 | |
| 0258+435 | 231 | 289 | 58 | 2.0 | >2.5 | | | 0.67 | 0.065 | | 2.7 | 86 | |
| 0300+162 | 134 | 110 | 24 | 2.0 | <2.5 | I | 14.5 | 0.77 | 0.032 | 25.45 | 1.8 | 17 | 2 |
| 0305+039 | 144 | 56 | 88 | 2.0 | <2.5 | I | 13.0 | 0.43 | 0.029 | 25.68 | 1.2 | 20 | 7 |
| 0307-305 | 78 | 93 | 15 | 3.4 | <2.5 | II | 16.5 | 0.54 | 0.068 | 25.15 | 2.4 | 26 | 7 |
| 0312-343 | 132 | 114 | 18 | 1.2 | <2.5 | | 15.6 | 0.62 | | | 2.0 | 26 | |
| 0314+412 | 57 | 32 | 25 | 3.0 | >2.5 | I | | | | | 2.6 | 58 | 7 |
| 0314+416 | 171 | 96 | 75 | 3.0 | >2.5 | I | 12.5 | 0.62 | 0.026 | 25.43 | 4.0 | 45 | 2 |
| 0320-374 | 60 | 126 | 66 | 3.8 | >2.5 | I | 8.9 | 0.52 | 0.005 | 25.46 | 2.6 | 26 | 7 |
| 0325+023 | 153 | 63 | 90 | 3.0 | >2.5 | II | 13.5 | 0.52 | 0.030 | 25.58 | 2.6 | 20 | 7 |
| 0326+396 | 128 | 82 | 46 | 1.0 | <2.5 | II | 14.9 | 0.60 | 0.024 | 24.68 | 2.4 | 15 | |
| 0331+391 | 101 | 180 | 79 | 1.0 | >2.5 | | 15.0 | 0.52 | 0.020 | 24.48 | 2.6 | 15 | |
| 0332-39 | 25 | 140 | 65 | 1.7 | >2.5 | | 15.3 | 1.05 | | | 3.5 | 26 | |
| 0336-355 | 112 | 51 | 61 | 1.2 | <2.5 | I | 10.9 | 0.80 | 0.005 | 23.52 | 2.4 | 26 | 7 |
| 0344-345 | 103 | 104 | 1 | 1.8 | <2.5 | I | 17.0 | 0.73 | 0.054 | 25.40 | 2.0 | 25 | 7 |
| 0349-279 | 72 | 53 | 19 | | >2.5 | II | 17.0 | 0.72 | 0.066 | 26.48 | 2.6 | 20 | 7 |
| 0349+212 | 126 | 17 | 71 | | >2.5 | | 16.0 | 0.70 | 0.133 | | 3.5 | 87 | |
| 0356+102 | 72 | 25 | 47 | 2.0 | >2.5 | II | 14.2 | 0.78 | 0.031 | 26.02 | 3.5 | 18 | 2 |
| 0427-539 | | 80 | 5 | | <2.5 | I | 13.2 | 0.70 | 0.038 | 25.55 | 2.4 | 54 | 7 |
| 0429-51 | 8 | 1 | 7 | 2.6 | <2.5 | | 12.7 | | | | 2.3 | 63 | |
| 0434-225 | 149 | 109 | 40 | 0.6 | <2.5 | I | 14.6 | 0.74 | 0.069 | 25.20 | 2.4 | 26 | 7 |
| 0446-208 | 174 | 18 | 24 | 0.6 | <2.5 | | 16.4 | 1.00 | | | 2.2 | 26 | |
| 0449-175 | 0 | 145 | 35 | 1.7 | <2.5 | I | 13.7 | 1.10 | 0.031 | 24.34 | 2.4 | 26 | 7 |
| 0452-190 | 140 | 82 | 58 | 3.1 | >2.5 | | 14.5 | 0.54 | | | 3.3 | 26 | |
| 0453-206 | 172 | 112 | 60 | 0.4 | <2.5 | I | 14.0 | 0.73 | 0.035 | 25.22 | 1.6 | 26 | 7 |
| 0511-305 | 85 | 33 | 52 | 1.3 | >2.5 | II | 17.0 | 0.84 | 0.058 | 25.39 | 2.7 | 20 | 3 |
| 0518-458 | 96 | 102 | 6 | 3.0 | <2.5 | II | 15.7 | 1.07 | 0.035 | 26.86 | 2.0 | 21 | 7 |
| 0521-365 | 75 | 123 | 48 | 2.6 | <2.5 | I | 15.3 | 0.43 | 0.061 | 26.64 | 1.4 | 26 | 16 |
| 0523-327 | 156 | 157 | 1 | 1.7 | >2.5 | II | 15.4 | 0.94 | 0.076 | 25.30 | 3.5 | 26 | 7 |
| 0546-329 | 175 | 8 | 13 | 1.8 | <2.5 | I | 14.5 | 0.97 | 0.037 | 24.73 | 2.2 | 26 | 7 |
| 0548-317 | 4 | 72 | 68 | 2.4 | >2.5 | II | 14.5 | 0.66 | 0.033 | 24.53 | 2.7 | 26 | 7 |
| 0632+263 | 16 | 115 | 81 | 0.1 | >2.5 | | 15.0 | | 0.040 | | 3.8 | 14 | |
| 0634-205 | 178 | 177 | 1 | 1.6 | >2.5 | I | 16.8 | 0.80 | 0.056 | 26.48 | 3.7 | 21 | 7 |
| 0651+542 | 129 | 102 | 27 | | >2.5 | II | 19.0 | 0.87 | 0.238 | 27.39 | 2.7 | 31 | 2 |
| 0652+426 | 124 | 50 | 74 | 2.0 | <2.5 | | | | | | 2.0 | 13 | |
| 0712-349 | 106 | 133 | 27 | 1.8 | <2.5 | | 15.9 | 0.55 | | | 2.0 | 26 | |
| 0712+534 | 120 | 114 | 6 | 1.0 | <2.5 | I | 15.0 | 0.60 | 0.064 | 24.83 | 2.2 | 13 | 15 |
| 0714+286 | 73 | 133 | 60 | 3.0 | >2.5 | | 16.0 | | 0.083 | | 2.6 | 13 | |
| 0718-340 | 56 | 63 | 7 | 0.9 | >2.5 | II | 16.5 | 0.50 | 0.030 | 24.71 | 2.9 | 26 | 7 |
| 0734+806 | 49 | 150 | 79 | | >2.5 | II | 17.0 | 0.68 | 0.118 | 26.68 | 3.1 | 17 | 2 |
| 0744+559 | 70 | 63 | 7 | 2.0 | <2.5 | II | 15.2 | 0.77 | 0.035 | 25.82 | 2.2 | 76 | 2 |
| 0745+521 | 37 | 92 | 55 | 1.0 | >2.5 | II | | 0.68 | 0.063 | | 3.0 | 83 | |
| 0755+379 | 144 | 107 | 37 | | <2.5 | I | 13.2 | 0.59 | 0.041 | 25.63 | 2.2 | 13 | |
| 0800+248 | 53 | 70 | 17 | | <2.5 | I | 15.7 | 0.68 | 0.043 | 24.41 | 2.3 | 15 | 7 |
| 0802+243 | 13 | 118 | 75 | 0.1 | >2.5 | II | 15.2 | 0.79 | 0.060 | 26.24 | 3.0 | 18 | 2 |
| 0810+66 | | 85 | 60 | | <2.5 | | 15.7 | | | | 1.5 | 57 | |
| 0818+472 | 103 | 4 | 81 | 1.0 | >2.5 | II | 16.5 | 0.69 | 0.130 | | 2.6 | 45 | 4 |
| 0819-30 | 44 | 119 | 75 | | >2.5 | II | 18.0 | 0.68 | 0.086 | | 3.1 | 20 | |
| 0819+061 | 98 | 38 | 60 | | >2.5 | II | 18.0 | 0.69 | 0.082 | 26.23 | 2.7 | 20 | 7 |
| 0836+299 | 59 | 30 | 29 | 2.0 | <2.5 | I | 15.7 | 0.78 | 0.065 | 25.68 | 1.8 | 15 | 15 |
| 0843+316 | 42 | 45 | 3 | | >2.5 | | 16.5 | 0.85 | 0.068 | 25.86 | 2.8 | 59 | |
| 0844+540 | 45 | 113 | 68 | 1.0 | >2.5 | | 15.0 | | 0.045 | | 2.9 | 85 | |
| 0844+319 | 123 | 170 | 47 | 1.0 | <2.5 | I | 13.5 | 0.78 | 0.068 | 25.86 | 2.4 | 15 | |
| 0908+376 | 80 | 5 | 75 | | >2.5 | II | 15.6 | 0.56 | 0.105 | 25.73 | 2.6 | 62 | |
| 0913+385 | 30 | 42 | 12 | | <2.5 | | 15.7 | 0.82 | 0.071 | 26.24 | 1.5 | 59 | |
| 0915+320 | 46 | 31 | 15 | 1.0 | <2.5 | I | 15.5 | 0.48 | 0.062 | 24.88 | 1.8 | 15 | 7 |
| 0915-119 | 130 | 24 | 74 | | >2.5 | I | | | 0.065 | | 2.6 | 64 | 7 |
| 0916+342 | 30 | 110 | 80 | | >2.5 | | 13.0 | 0.87 | 0.017 | 23.60 | 3.2 | 15 | |
| 0922+366 | 130 | 170 | 40 | | <2.5 | I | 15.5 | 0.98 | 0.112 | 25.99 | 2.0 | 16 | |
| 0923+330 | | | 5 | | <2.5 | | 16.0 | 1.12 | 0.140 | | 1.9 | 16 | |
| 0924+302 | 49 | 55 | 6 | 1.0 | <2.5 | | 14.5 | 1.04 | 0.027 | 24.72 | 2.0 | 72 | |
| 0936+361 | 118 | 164 | 46 | | >2.5 | II | 16.8 | 0.74 | 0.137 | | 6.0 | 18 | 2 |
| 0938+399 | 45 | 14 | 31 | 3.0 | >2.5 | II | 16.2 | 0.56 | 0.108 | 26.31 | 4.0 | 16 | 4 |
| 0940-304 | 90 | 21 | 69 | 5.0 | <2.5 | | 14.5 | 0.58 | | | 1.5 | 26 | |
| 1000+201 | 112 | 7 | 75 | | >2.5 | I | 16.5 | 0.80 | 0.168 | 26.56 | 2.6 | 85 | 7 |
| 1002-320 | 52 | 29 | 23 | 1.9 | >2.5 | | 17.4 | 0.93 | | | 2.8 | 26 | |
| 1003+351 | 45 | 123 | 78 | 3.0 | >2.5 | II | 15.5 | 0.51 | 0.099 | 26.62 | 3.5 | 73 | 2 |
| 1005+007 | 38 | 71 | 33 | 1.0 | <2.5 | | 15.4 | | | | 2.4 | 38 | |
| 1005+282 | 5 | 150 | 45 | 2.0 | >2.5 | II | 16.4 | 1.15 | 0.148 | 25.36 | 2.6 | 59 | |
| 1014+398 | 115 | 130 | 15 | | >2.5 | II | 15.5 | 1.10 | 0.106 | | 5.0 | 16 | |
| 1015+491 | 95 | 10 | 85 | | >2.5 | I | 14.8 | 0.57 | 0.080 | | 3.2 | 62 | |
| 1033+003 | 131 | 8 | 57 | 2.0 | <2.5 | | 15.2 | | | | 1.8 | 85 | |
| 1040+317 | | 50 | 21 | | <2.5 | I | 15.5 | 0.62 | 0.036 | 24.97 | 2.0 | 15 | 7 |
| 1053-282 | 48 | 26 | 22 | 3.3 | >2.5 | II | 15.5 | 0.61 | 0.061 | 25.30 | 2.7 | 26 | 7 |
| 1102+304 | 147 | 70 | 77 | 2.0 | >2.5 | II | 15.7 | 0.72 | 0.072 | 25.32 | 3.8 | 15 | |
| 1107-372 | 30 | 78 | 48 | 2.3 | <2.5 | | 12.4 | 0.70 | | 22.80 | 1.8 | 26 | |
| 1108+272 | | 80 | 5 | | <2.5 | I | 14.6 | 0.48 | 0.033 | 23.01 | 2.3 | 15 | 7 |
| 1113+295 | 138 | 71 | 67 | 2.0 | >2.5 | II | 14.2 | 0.64 | 0.049 | 25.70 | 2.8 | 15 | 7 |
| 1116+281 | 40 | 113 | 73 | | >2.5 | | 14.3 | 0.65 | 0.067 | 25.30 | 2.7 | 59 | |
| 1122+390 | 35 | 118 | 83 | 2.1 | >2.5 | I | 11.6 | 0.57 | 0.007 | 23.98 | 2.9 | 28 | 7 |
| 1123-351 | 174 | 120 | 54 | 1.8 | <2.5 | | 16.0 | 0.70 | 0.033 | | 2.2 | 26 | |
| 1127+012 | 100 | 12 | 88 | 3.0 | >2.5 | | 16.7 | | | | 2.7 | 85 | |
| 1137+123 | 139 | 12 | 53 | 2.0 | <2.5 | | 16.5 | | | | 1.6 | 85 | |
| 1141+374 | 130 | 52 | 78 | | >2.5 | II | 15.9 | 0.94 | 0.115 | 26.46 | >5 | 23 | |
| 1141+466 | 147 | 40 | 73 | | >2.5 | II | 15.8 | 1.10 | 0.162 | | 2.6 | 23 | |
| 1142-341 | 31 | 150 | 61 | 2.1 | >2.5 | | 15.6 | 0.92 | | | 2.8 | 26 | |
| 1146-11 | 79 | 104 | 25 | | <2.5 | II | 18.0 | 0.96 | 0.117 | | 1.3 | 21 | |
| 1154-038 | 45 | 109 | 64 | 2.0 | >2.5 | | 14.3 | | | | 3.3 | 85 | |
| 1155+266 | 55 | 130 | 75 | | >2.5 | | | | | | 2.7 | 56 | |
| 1204+241 | | 166 | 5 | | <2.5 | | 15.2 | 0.76 | 0.077 | 24.83 | 1.5 | 59 | |
| 1209+746 | 60 | 155 | 85 | | >2.5 | | 16.5 | | 0.061 | | 3.5 | 61 | |
| 1216+061 | 150 | 83 | 67 | 3.0 | >2.5 | II | 11.0 | 0.51 | 0.007 | 24.80 | 3.0 | 20 | 7 |
| 1218+296 | 40 | 152 | 68 | 0.7 | <2.5 | | 11.2 | 0.24 | 0.002 | 21.60 | 1.8 | 65 | |
| 1222+131 | 116 | 167 | 51 | 1.0 | <2.5 | I | 10.0 | 0.60 | 0.003 | 23.80 | 1.9 | 17 | 2 |
| 1225+265 | 50 | 70 | 20 | | <2.5 | II | 16.1 | 0.79 | 0.064 | | 2.4 | 59 | |
| 1227+83 | 160 | 70 | 90 | 1.8 | <2.5 | | 12.8 | | | | 1.5 | 66 | |
| 1228-335 | 164 | 83 | 81 | 2.0 | <2.5 | | 15.4 | 0.60 | | | 2.4 | 26 | |
| 1228+127 | 157 | 101 | 56 | 1.4 | <2.5 | I | 8.7 | 0.79 | 0.004 | 25.65 | 2.0 | 17 | 2 |
| 1240+029 | 166 | 33 | 47 | 1.9 | >2.5 | | 12.9 | | | | 2.6 | 63 | |
| 1249+035 | 27 | 146 | 61 | 2.0 | >2.5 | | | | | | 2.6 | 85 | |
| 1250-102 | 65 | 162 | 83 | 1.0 | >2.5 | | 12.0 | 1.20 | 0.014 | 23.27 | 4.0 | 37 | |
| 1251+278 | 30 | 169 | 41 | 0.1 | <2.5 | I | 15.5 | 0.58 | 0.086 | 26.27 | 1.5 | 19 | 15 |
| 1254+277 | 51 | 11 | 40 | 3.0 | <2.5 | I | 12.3 | 0.86 | 0.025 | 22.63 | 1.8 | 15 | 7 |
| 1256+281 | 171 | 275 | 76 | 2.0 | >2.5 | I | 14.9 | 1.04 | 0.024 | 24.50 | 2.6 | 74 | 15 |
| 1257-253 | 37 | 150 | 67 | 2.4 | >2.5 | | 16.0 | 0.70 | | | 3.5 | 26 | |
| 1257+282 | 17 | 39 | 22 | | <2.5 | I | 14.0 | 0.75 | 0.023 | 23.05 | 2.2 | 67 | |
| 1258-321 | 167 | 125 | 42 | 3.2 | <2.5 | | 12.8 | 0.59 | | | 1.8 | 26 | |
| 1313+072 | 40 | 71 | 31 | | <2.5 | | 15.5 | | 0.051 | 24.75 | 2.0 | 20 | |
| 1316+299 | 67 | 97 | 30 | | <2.5 | | 15.0 | 0.71 | 0.073 | 25.85 | 1.5 | 13 | |
| 1317+258 | 75 | 54 | 21 | 2.0 | >2.5 | | | | | | 2.6 | 86 | |
| 1318-434 | 100 | 24 | 76 | 2.0 | >2.5 | I | 14.5 | 0.96 | 0.011 | 25.01 | 3.9 | 21 | 7 |
| 1319+428 | 130 | 79 | 51 | | >2.5 | II | 16.0 | 0.95 | 0.079 | 26.15 | 3.0 | 17 | 2 |
| 1321+318 | 69 | 111 | 42 | | >2.5 | I | 13.9 | 0.65 | 0.016 | 24.60 | 3.0 | 67 | 15 |
| 1322-428 | | 40 | 10 | 2.0 | <2.5 | I | 7.0 | 0.79 | 0.002 | 23.80 | 2.4 | 54 | 8 |
| 1322+366 | 75 | 7 | 68 | 3.0 | >2.5 | II | 14.0 | 0.46 | 0.018 | 24.35 | 3.5 | 13 | |
| 1323-271 | 154 | 68 | 86 | 3.9 | >2.5 | II | 12.9 | 0.67 | 0.043 | 24.99 | 4.0 | 26 | |
| 1323+370 | 87 | 154 | 67 | 2.0 | >2.5 | II | 15.0 | 0.70 | 0.080 | | 2.7 | 62 | |
| 1331-099 | 55 | 107 | 52 | | >2.5 | II | 17.5 | 0.90 | 0.081 | 26 26 | 2.8 | 21 | 16 |
| 1333-337 | 47 | 125 | 78 | 0.9 | >2.5 | II | 11.9 | 0.79 | 0.013 | 25.06 | 3.3 | 26 | 17 |
| 1344-241 | 149 | 155 | 6 | 3.9 | <2.5 | | 14.4 | 1.05 | | | 2.0 | 26 | |
| 1346+268 | 18 | 20 | 2 | | <2.5 | I | 13.8 | 0.92 | 0.063 | 25.73 | 2.0 | 13 | 7 |
| 1350+316 | 62 | 100 | 38 | | <2.5 | I | 15.6 | 0.70 | 0.045 | 25.40 | 1.5 | 19 | 2 |
| 1354-251 | 147 | 155 | 8 | 3.9 | <2.5 | | 15.4 | 0.64 | | | 1.5 | 26 | |
| 1357+287 | 100 | 15 | 85 | | >2.5 | II | 14,8 | 0.80 | 0.063 | 25.06 | 2.6 | 59 | |
| 1358-113 | 41 | 125 | 84 | 2.0 | >2.5 | II | 15.0 | 0.70 | 0.037 | 24.89 | 3.0 | 29 | |
| 1400-337 | 90 | 4 | 86 | 2.1 | <2.5 | | 12.4 | 1.28 | 0.014 | | 1.4 | 26 | |
| 1401+35 | 70 | 0 | 70 | | <2.5 | | 12.8 | 0.92 | 0.013 | 23.60 | 1.5 | 62 | |
| 1401-05 | | | 15 | | <2.5 | | 17.0 | | | | 1.9 | 21 | |
| 1407+177 | 6 | 73 | 67 | 2.0 | <2.5 | I | 13.4 | | 0.016 | 23.68 | 1.8 | 21 | 15 |
| 1411+094 | 84 | 178 | 86 | | >2.5 | | 18.3 | | 0.162 | | 2.8 | 85 | |
| 1413-36 | 43 | 35 | 8 | 2.6 | >2.5 | | 17.5 | 0.74 | | | 3.0 | 26 | |
| 1414+110 | 146 | 85 | 61 | 1.0 | <2.5 | I | 13.3 | 0.67 | 0.024 | 25.36 | 1.6 | 20 | 2 |
| 1420+198 | 95 | 135 | 40 | | >2.5 | II | 18.0 | 0.78 | 0.270 | 27.58 | 3.0 | 45 | 2 |
| 1422+268 | 118 | 96 | 22 | 2.0 | <2.5 | I | 15.6 | 0.74 | 0.037 | 25.07 | 2.4 | 15 | 7 |
| 1427+07 | 54 | 157 | 77 | | >2.5 | | 15.6 | | | | 2.6 | 20 | |
| 1433+553 | 110 | 143 | 33 | 1.0 | <2.5 | | 17.0 | 0.80 | 0.140 | 25.05 | 2.3 | 87 | |
| 1441+262 | 150 | 68 | 82 | | >2.5 | II | 14.3 | 0.79 | 0.062 | 25.03 | 3.1 | 59 | |
| 1441+522 | | 125 | 25 | | <2.5 | II | 17.0 | 0.76 | 0.141 | 26.76 | 2.2 | 19 | 2 |
| 1449-129 | 135 | 89 | 46 | | <2.5 | I | 18.0 | | 0.070 | 25.26 | 2.2 | 20 | 7 |
| 1452+165 | 161 | 59 | 78 | 3.0 | <2.5 | II | 14.9 | 0.71 | 0.046 | | 1.5 | 21 | |
| 1457+29 | 130 | 169 | 39 | | <2.5 | | 17.2 | | | | 1.3 | 59 | |
| 1458+21A | | | 0 | | <2.5 | | | | | | 2.2 | 56 | |
| 1459+21E | | | 20 | | <2.5 | | | | | | 2.3 | 56 | |
| 1502+262 | | 150 | 12 | | >2.5 | I | 15.2 | 0.92 | 0.054 | 26.46 | 3.3 | 13 | 2 |
| 1509+059 | 26 | 145 | 61 | 3.3 | >2.5 | | | | | | 2.7 | 63 | |
| 1512+30 | 110 | 31 | 79 | | >2.5 | II | 15.4 | 0.75 | 0.093 | 24.99 | 2.8 | 59 | |
| 1514+004 | 53 | 132 | 79 | 3.0 | >2.5 | II | 16.5 | 0.40 | 0.052 | | 3.1 | 20 | |
| 1514+072 | 21 | 16 | 5 | 3.0 | <2.5 | I | 16.0 | 1.02 | 0.035 | 26.10 | 1.4 | 29 | 4 |
| 1519+078 | | | 90 | | >2.5 | | 15.0 | 1.93 | 0.046 | 25.09 | 2.6 | 57 | |
| 1525+291 | 8 | 14 | 6 | | <2.5 | I | 15.4 | 0.73 | 0.065 | 24.89 | 1.5 | 15 | 7 |
| 1527+308 | 175 | 130 | 45 | | >2.5 | | 15.0 | 0.98 | 0.114 | 25.39 | 2.7 | 59 | |
| 1547+309 | 136 | 120 | 16 | | >2.5 | | 16.5 | 0.96 | 0.111 | | 3.2 | 37 | |
| 1549+202 | 119 | 80 | 39 | | >2.5 | II | 18.5 | 0.88 | 0.090 | 26.5 | 3.8 | 18 | 2 |
| 1553+245 | 19 | 129 | 70 | 3.0 | >2.5 | I | 14.4 | 0.28 | 0.043 | 23.36 | 3.5 | 13 | 7 |
| 1555+308 | 120 | 122 | 2 | | >2.5 | | 16.1 | 0.58 | 0.075 | | 3.2 | 59 | |
| 1556+274 | 109 | 291 | 2 | 2.0 | <2.5 | | | | | | 2.4 | 74 | |
| 1559+021 | 138 | 100 | 38 | 3.0 | >2.5 | II | 15.5 | 0.61 | 0.104 | 27.00 | 2.9 | 20 | 4 |
| 1601+173 | 64 | 180 | 64 | 2.0 | <2.5 | | 13.5 | | 0.034 | | 1.8 | 91 | |
| 1602+178 | 117 | 171 | 54 | 1.0 | <2.5 | I | 14.6 | 0.15 | 0.032 | | 2.0 | 85 | 7 |
| 1602+34 | | | 37 | | <2.5 | | 15.4 | 0.82 | 0.032 | 23.4 | 2.4 | 15 | |
| 1604+183 | 89 | 176 | 87 | 2.0 | >2.5 | | 15.0 | | | | 2.8 | 90 | |
| 1610+296 | 1 | 66 | 65 | 3.0 | <2.5 | I | 14.8 | 0.72 | 0.031 | 24.13 | 1.5 | 15 | 7 |
| 1610-607 | 127 | 86 | 41 | 2.0 | >2.5 | II | 12.8 | 1.15 | 0.017 | | 3.4 | 54 | |
| 1615+325 | 28 | 17 | 11 | | >2.5 | II | 16.0 | 0.61 | 0.152 | 26.69 | 3.0 | 19 | 7 |
| 1615+351 | 252 | 323 | 71 | 1.0 | >2.5 | II | 14.9 | 0.76 | 0.030 | 25.31 | 3.5 | 60 | 7 |
| 1621+380 | 175 | 70 | 75 | 5.0 | >2.5 | I | 14.1 | 0.56 | 0.031 | 24.58 | 2.8 | 60 | 7 |
| 1626+397 | 34 | 82 | 48 | | <2.5 | I | 12.0 | 1.19 | 0.030 | 25.87 | 2.0 | 17 | 2 |
| 1636+379 | | 75 | 5 | 1.0 | <2.5 | I | 16.4 | 0.80 | 0.179 | | 2.3 | 56 | |
| 1637-771 | 89 | 165 | 76 | 4.0 | >2.5 | II | 16.0 | 0.50 | 0.043 | | 2.6 | 25 | 8 |
| 1640+826 | 27 | 124 | 83 | 2.0 | >2.5 | I | 14.0 | | | | 3.0 | 76 | |
| 1643+274 | 120 | 35 | 85 | | >2.5 | II | 15.8 | 0.92 | 0.102 | 25.10 | 2.7 | 59 | |
| 1648+050 | 127 | 100 | 27 | | <2.5 | I | 19.0 | 1.00 | 0.154 | 28.26 | 2.2 | 20 | 7 |
| 1652+39 | 170 | 135 | 35 | | <2.5 | | 13.7 | 0.18 | 0.034 | 24.35 | 1.8 | 81 | |
| 1655+32A | | | 50 | | <2.5 | | | | | | 2.1 | 56 | |
| 1657+325 | | 10 | 5 | | <2.5 | | 16.8 | 0.75 | 0.063 | 25.32 | 1.5 | 56 | |
| 1658+302 | | 70 | 10 | | <2.5 | I | 14.7 | 0.66 | 0.035 | 24.91 | 1.5 | 16 | 15 |
| 1658+326 | 24 | 10 | 14 | 2.0 | <2.5 | | 16.1 | 0.85 | 0.102 | 25.42 | 1.7 | 56 | |
| 1658+32B | | | 68 | | >2.5 | | | | | | 2.6 | 56 | |
| 1710+156 | 5 | 169 | 16 | 4.0 | <2.5 | | 16.7 | | | | 1.5 | 85 | |
| 1712+641 | 208 | 150 | 58 | 1.0 | >2.5 | | 17.3 | 0.74 | 0.081 | | 2.8 | 57 | |
| 1717-009 | 35 | 89 | 54 | | >2.5 | II | 16.8 | 0.71 | 0.030 | 26.70 | 2.6 | 20 | 4 |
| 1726+318 | 78 | 110 | 32 | | >2.5 | II | 15.5 | 0.57 | 0.166 | 26.85 | 2.9 | 16 | 11 |
| 1741+390 | | 90 | 75 | | >2.5 | | 15.0 | | 0.042 | | 2.6 | 16 | |
| 1744+557 | 10 | 77 | 67 | 2.0 | >2.5 | | 13.2 | | 0.030 | | 2.8 | 13 | |
| 1747+303 | 70 | 150 | 80 | | >2.5 | | 16.7 | 1.17 | 0.130 | 23.96 | 2.9 | 59 | |
| 1752+325 | 110 | 41 | 69 | | >2.5 | | 14.3 | 0.91 | 0.045 | 24.44 | 2.6 | 59 | |
| 1759+211 | 60 | 50 | 10 | 2.0 | <2.5 | | 17.5 | | | | 2.3 | 89 | |
| 1820+689 | 133 | 177 | 44 | 2.0 | <2.5 | | 15.0 | 0.70 | 0.131 | | 1.5 | 88 | |
| 1826+743 | 147 | 161 | 14 | | >2.5 | II | 18.0 | 0.68 | 0.256 | | 2.8 | 17 | 11 |
| 1833+326 | 73 | 48 | 25 | 2.0 | >2.5 | II | 14.5 | 0.59 | 0.058 | 26.30 | 2.7 | 17 | 2 |
| 1833+653 | 97 | 19 | 78 | | >2.5 | | 17.0 | | 0.161 | | 2.6 | 85 | |
| 1834+197 | 22 | 142 | 60 | 1.0 | >2.5 | | 14.0 | 0.79 | 0.016 | | 2.7 | 13 | |
| 1842+455 | 51 | 68 | 17 | | >2.5 | II | 15.0 | 0.70 | 0.091 | 25.73 | 3.2 | 19 | 2 |
| 1845+797 | 60 | 145 | 85 | | >2.5 | II | 14.4 | 0.75 | 0.056 | 26.56 | 5.0 | 17 | 2 |
| 1855+379 | 55 | 4 | 51 | | <2.5 | I | 14.9 | 0.84 | 0.055 | 25.02 | 1.1 | 15 | 7 |
| 1928-340 | 138 | 9 | 51 | 1.3 | >2.5 | II | 17.0 | 0.70 | 0.098 | 26.21 | 3.0 | 26 | 16 |
| 1929-397 | 130 | 124 | 6 | 1.2 | <2.5 | | 16.0 | 0.70 | 0.075 | | 2.4 | 26 | |
| 1939+606 | 8 | 26 | 18 | | <2.5 | II | 18.0 | 0.71 | 0.201 | 27.40 | 1.9 | 19 | 2 |
| 1940+504 | 34 | 28 | 6 | 1.0 | <2.5 | I | 14.0 | 0.56 | 0.024 | 25.23 | 1.5 | 17 | 11 |
| 1949+023 | 163 | 92 | 71 | | >2.5 | II | 15.0 | 0.45 | 0.059 | 26.33 | 2.6 | 20 | 4 |
| 1957+405 | 152 | 109 | 43 | | >2.5 | II | 15.0 | 0.74 | 0.057 | 25.76 | 2.9 | 17 | 4 |
| 2013-308 | 123 | 64 | 59 | 1.9 | >2.5 | I | 15.4 | 0.86 | 0.089 | 25.33 | 2.8 | 26 | 7 |
| 2014-558 | 11 | 157 | 34 | | <2.5 | | 15.5 | 0.70 | 0.061 | | 2.2 | 21 | |
| 2031-359 | 146 | 170 | 24 | 1.1 | <2.5 | | 15.5 | 0.78 | | | 1.5 | 26 | |
| 2040-267 | 68 | 158 | 90 | 0.1 | >2.5 | II | 13.5 | 0.73 | 0.038 | 24.98 | 3.4 | 20 | 7 |
| 2053-201 | 11 | 52 | 41 | | <2.5 | I | 17.8 | | 0.156 | 26.29 | 2.4 | 92 | 7 |
| 2058-135 | 29 | 101 | 72 | 1.0 | <2.5 | II | 15.5 | 0.81 | 0.046 | 24.89 | 2.0 | 21 | |
| 2058-282 | 55 | 135 | 80 | 0.8 | >2.5 | I | 14.8 | 0.74 | 0.038 | 25.67 | 3.0 | 20 | 7 |
| 2059-311 | 24 | 106 | 82 | 3.7 | >2.5 | | 14.5 | 0.50 | | | 3.5 | 26 | |
| 2103+124 | 59 | 138 | 79 | 2.0 | >2.5 | | 17.3 | 0.56 | | | 3.0 | 85 | |
| 2104-256 | 138 | 22 | 64 | | >2.5 | II | 16.8 | 0.89 | 0.039 | 25.30 | 3.2 | 26 | 8 |
| 2116+262 | 65 | 22 | 43 | 5.0 | <2.5 | I | 14.0 | | 0.016 | 23.57 | 1.8 | 15 | 7 |
| 2117+605 | 106 | 35 | 71 | 2.0 | >2.5 | II | 15.0 | 0.72 | 0.054 | | 2.8 | 19 | 4 |
| 2121+248 | | 4 | 85 | | <2.5 | I | 15.5 | 0.75 | 0.102 | 27.09 | 2.0 | 18 | 2 |
| 2128-388 | 106 | 49 | 57 | 1.3 | >2.5 | | 14.4 | 0.64 | | | 2.7 | 26 | |
| 2141+279 | 35 | 173 | 42 | | >2.5 | II | 18.5 | 0.86 | 0.215 | 27.31 | 2.9 | 17 | 2 |
| 2152-699 | 130 | 14 | 64 | | >2.5 | I | 13.8 | 0.71 | 0.027 | 26.38 | 2.8 | 54 | 7 |
| 2158-380 | 97 | 50 | 47 | 2.7 | >2.5 | | 14.6 | 0.71 | | | 2.8 | 26 | |
| 2225-308 | 145 | 141 | 4 | 0.3 | >2.5 | I | 15.8 | 0.74 | 0.055 | 24.90 | 4.5 | 26 | 7 |
| 2229+391 | 7 | 9 | 2 | 2.0 | <2.5 | I | 13.0 | 0.58 | 0.017 | 24.93 | 2.2 | 18 | 2 |
| 2236-176 | 94 | 53 | 41 | 1.3 | >2.5 | I | 15.3 | 0.81 | 0.075 | 25.38 | 3.5 | 26 | 7 |
| 2236-364 | 49 | 132 | 83 | 4.6 | <2.5 | | 15.2 | 0.57 | | | 1.8 | 26 | |
| 2236+350 | 5 | 46 | 41 | | <2.5 | I | 15.0 | 0.58 | 0.028 | 24.40 | 2.4 | 15 | 15 |
| 2244+366 | 131 | 34 | 83 | | <2.5 | II | 16.0 | 0.80 | 0.082 | | 2.2 | 16 | |
| 2247+113 | 48 | 31 | 17 | 1.0 | <2.5 | I | 14.4 | 0.75 | 0.023 | 25.21 | 2.0 | 21 | 2 |
| 2318+079 | 2 | 30 | 28 | 1.7 | >2.5 | I | 12.8 | | 0.011 | 23.17 | 2.6 | 67 | 15 |
| 2333-327 | 88 | 132 | 44 | 1.9 | <2.5 | | 14.6 | 0.61 | | | 1.5 | 26 | |
| 2335+267 | 60 | 140 | 80 | | >2.5 | I | 13.2 | 0.75 | 0.029 | 25.88 | 4.0 | 17 | 2 |
| 2350-374 | 25 | 56 | 31 | 2.4 | >2.5 | | 16.0 | 0.55 | | | 3.0 | 26 | |
| 2353-184 | 153 | 140 | 13 | 0.9 | <2.5 | | 16.0 | 0.78 | | | 1.3 | 26 | |
| 2353+56 | 135 | 115 | 20 | 4.0 | >2.5 | | | | | | 6.0 | 63 | |
| 2354-351 | 162 | 150 | 12 | 2.4 | >2.5 | | 14.4 | 1.20 | 0.049 | | 3.0 | 26 | |
| 2354+471 | 52 | 64 | 12 | 1.0 | <2.5 | I | 15.0 | 0.72 | 0.046 | 24.63 | 2.3 | 28 | 15 |
| 2356-611 | 3 | 134 | 49 | 2.0 | >2.5 | II | 16.0 | 1.36 | 0.096 | 27.79 | 3.0 | 54 | 3 |

Sample 2.

Col.1,2: Radio Source Name

Col.3: dPA is a relative position angle between radio axis and integrated polarisation;

Col.4: Ref. for the data of dPA

Col.5: K is the ratio of major to minor axis of radio image

Col.6: Ref. for the radio maps

Col.7: FR is the Fanaroff Riley classe

Col.8: Ref. for the FR classe

| 1<br>Source | 2<br>Name | 3<br>dPA | 4<br>Samp | 5<br>K | 6<br>Ref | 7<br>FR | 8<br>ref |
|---|---|---|---|---|---|---|---|
| 0002+12 | | 73 | Cl | 3.5 | 20 | | |
| 0003-00 | 3c2 | 39 | Ha | 2.2 | 30 | | |
| 0007+12 | 4c12.03 | 83 | Cl | 2.6 | 20 | II | 5 |
| 0013+79 | 3c6.1 | 64 | PB | 3.5 | 19 | II | 2 |
| 0017+15 | 3c9 | 71 | Cl | 2.2 | 24 | II | 2 |
| 0020-25 | | 79 | Cl | 1.7 | 20 | | |
| 0031+39 | 3c13 | 69 | Cl | 5.0 | 27 | II | 2 |
| 0033+18 | 3c14 | 89 | Cl | 3.0 | 27 | II | 2 |
| 0034-01 | 3c15 | 85 | Cl | 2.6 | 27 | II | 1 |
| 0035+38 | 4c38.03 | 84 | PB | 3.6 | 28 | | |
| 0038+32 | 3c19 | 88 | Cl | 2.8 | 27 | II | 2 |
| 0040+51 | 3c20 | 44 | Cl | 4.0 | 19 | II | 2 |
| 0043-42 | | 0 | PB | 2.4 | 25 | II | 7 |
| 0048+50 | 3c22 | 79 | Cl | 5.0 | 27 | II | 2 |
| 0052+68 | 3c27 | 53 | Cl | 3.5 | 50 | II | 50 |
| 0104+32 | 3c31 | 51 | Cl | 2.0 | 27 | I | 2 |
| 0105+72 | 3c33.1 | 87 | Cl | 3.8 | 17 | II | 2 |
| 0106+13 | 3c33 | 73 | Cl | 3.4 | 17 | II | 2 |
| 0107+31 | 3c34 | 74 | Cl | 3.9 | 27 | II | 2 |
| 0114-47 | | 32 | Cl | 2.2 | 21 | II | 3 |
| 0115+02 | 3c37 | 70 | Ha | 3.0 | 32 | | |
| 0123+32 | 3c41 | 74 | Cl | 3.5 | 55 | II | 2 |
| 0125+28 | 3c42 | 60 | PB | 4.0 | 18 | II | 2 |
| 0128+25 | 4c25.07 | 22 | PB | 1.4 | 46 | | |
| 0128+06 | 3c44 | 89 | Cl | 2.7 | 18 | | |
| 0131-36 | | 15 | Ha | 2.6 | 20 | II | 7 |
| 0132+37 | 3c46 | 87 | Cl | 5.0 | 17 | II | 2 |
| 0133+20 | 3c47 | 6 | Cl | 2.2 | 17 | II | 2 |
| 0134+32 | 3c48 | 43 | Mi | 1.5 | 33 | | |
| 0145+53 | 3c52 | 50 | Cl | 2.1 | 17 | | |
| 0152+43 | 3c54 | 43 | Cl | 4.5 | 55 | | |
| 0154+28 | 3c55 | 83 | Cl | 3.7 | 24 | II | 2 |
| 0159-11 | 3c57 | 65 | Ha | 1.6 | 32 | | |
| 0210+86 | 3c61.1 | 38 | PB | 4.5 | 17 | II | 2 |
| 0211+34 | 4c34.06 | 58 | PB | 3.5 | 22 | | |
| 0213-13 | 3c62 | 20 | Cl | 2.7 | 80 | II | 7 |
| 0214-48 | | 84 | Cl | 3.3 | 54 | I | 7 |
| 0219+08 | 3c64 | 73 | Cl | 2.4 | 20 | | |
| 0220+39 | 3c65 | 69 | Cl | 3.5 | 55 | II | 2 |
| 0221+27 | 3c67 | 64 | Cl | 5.0 | 33 | II | 2 |
| 0222-00 | 4c-00.12 | 59 | Cl | 3.7 | 30 | | |
| 0229+34 | 3c68.1 | 88 | Cl | 2.7 | 18 | II | 2 |
| 0229+35 | | 79 | PB | 3.0 | 16 | | |
| 0232-02 | 4c-02.12 | 58 | Cl | 4.0 | 30 | | |
| 0234+58 | 3c69 | 88 | Cl | 4.5 | 19 | | |
| 0241-51 | | 72 | Cl | 2.8 | 21 | | |
| 0241+29 | 4c29.08 | 64 | PB | 2.5 | 22 | | |
| 0300+16 | 3c76.1 | 3 | Cl | 1.8 | 17 | I | 2 |
| 0307+16 | 3c79 | 78 | Cl | 4.5 | 29 | II | 2 |
| 0313+34 | 4c34.13 | 80 | Cl | 2.8 | 47 | | |
| 0323+55 | 3c86 | 55 | Cl | 5.0 | 17 | | |
| 0325+02 | 3c88 | 16 | Cl | 2.6 | 20 | I | 7 |
| 0336-35 | | 76 | Cl | 2.4 | 26 | | |
| 0344-34 | | 56 | Cl | 2.0 | 25 | I | 7 |
| 0349-27 | OE-283 | 77 | PB | 2.6 | 20 | II | 7 |
| 0349+26 | 4c26.12 | 77 | Cl | 4.0 | 22 | | |
| 0356+10 | 3c98 | 19 | Cl | 3.5 | 18 | II | 2 |
| 0403-13 | OF-105 | 85 | Ha | 2.3 | 48 | | |
| 0404+03 | 3c105 | 80 | Cl | 3.5 | 20 | II | 1 |
| 0404+42 | 3c103 | 86 | Cl | 3.2 | 17 | | |
| 0410+11 | 3c109 | 83 | Cl | 3.0 | 19 | II | 2 |
| 0415+37 | 3c111 | 81 | PB | 4.0 | 18 | II | 1 |
| 0427-53 | | 84 | Cl | 2.9 | 54 | | |
| 0431-133 | | 82 | PB | 2.8 | 25 | | |
| 0433+29 | 3c123 | 58 | Cl | 2.0 | 18 | II | 2 |
| 0453+22 | 3c132 | 66 | Cl | 1.9 | 18 | II | 2 |
| 0459+25 | 3c133 | 33 | Cl | 2.4 | 18 | II | 2 |
| 0501+38 | 3c134 | 87 | Cl | 3.4 | 17 | | |
| 0511-48 | | 14 | Cl | 1.4 | 21 | | |
| 0511-30 | | 46 | Cl | 2.7 | 20 | II | 3 |
| 0511+00 | 3c135 | 1 | Cl | 1.5 | 21 | II | 1 |
| 0515+50 | 3c137 | 84 | Cl | 5.0 | 53 | | |
| 0518-45 | PikA | 9 | Cl | 2.0 | 21 | II | 8 |
| 0518+16 | 3c138 | 80 | Ha | 2.2 | 33 | | |
| 0521+28 | 3c139.2 | 65 | Cl | 3.5 | 44 | | |
| 0528+06 | 3c142.1 | 72 | Cl | 4.2 | 79 | | |
| 0538+49 | 3c147 | 50 | Mi | 1.7 | 33 | | |
| 0605+48 | 3c153 | 20 | Cl | 1.7 | 24 | II | 2 |
| 0610+26 | 3c154 | 86 | Cl | 5.0 | 45 | | |
| 0618-37 | | 18 | Cl | 2.0 | 26 | | |
| 0634-20 | | 82 | PB | 3.7 | 21 | II | 14 |
| 0640+23 | 3c165 | 55 | Cl | 4.5 | 20 | | |
| 0651+54 | 3c171 | 16 | Cl | 2.7 | 31 | II | 2 |
| 0656-24 | | 81 | Cl | 1.1 | 20 | | |
| 0659+25 | 3c172 | 79 | Cl | 3.5 | 18 | II | 2 |
| 0702+74 | 3c173.1 | 32 | Cl | 2.2 | 17 | II | 2 |
| 0710+11 | 3c175 | 43 | Cl | 3.5 | 50 | II | 2 |
| 0711+14 | 3c175.1 | 68 | Cl | 2.4 | 27 | II | 2 |
| 0715-36 | | 59 | Cl | 1.7 | 21 | | |
| 0723+67 | 3c179 | 81 | Cl | 4.0 | 40 | II | 12 |
| 0724-01 | 3c180 | 18 | Cl | 2.1 | 20 | | |
| 0725+14 | 3c181 | 49 | Cl | 2.0 | 34 | II | 2 |
| 0733+70 | 3c184 | 64 | Cl | 5.0 | 24 | II | 2 |
| 0734+80 | 3c184.1 | 88 | Cl | 3.1 | 17 | II | 2 |
| 0736-06 | 01-161 | 69 | Cl | 3.0 | 21 | | |
| 0742+02 | 3c187 | 76 | Cl | 2.6 | 20 | | |
| 0755+37 | 4c37.21 | 2 | Cl | 2.2 | 13 | I | |
| 0800-09 | | 19 | Cl | 2.0 | 21 | | |
| 0802+10 | 3c191 | 34 | Ha | 2.3 | 34 | II | 2 |
| 0802+24 | 3c192 | 49 | Cl | 3.0 | 18 | II | 2 |
| 0809+48 | 3c196 | 47 | Cl | 1.5 | 24 | II | 2 |
| 0814+22 | 4c22.20 | 19 | Cl | 2.0 | 45 | | |
| 0818+47 | 3c197.1 | 22 | Cl | 2.6 | 45 | II | 6 |
| 0819-30 | | 55 | PB | 3.1 | 20 | II | |
| 0819+06 | 3c198 | 57 | Cl | 2.7 | 20 | II | 6 |
| 0824+29 | 3c200 | 57 | Cl | 3.5 | 27 | II | 2 |
| 0833+65 | 3c204 | 15 | PB | 1.7 | 18 | II | 2 |
| 0835+58 | 3c205 | 59 | Cl | 2.2 | 18 | II | 2 |
| 0836+19 | 4c19.31 | 76 | Cl | 3.0 | 27 | | |
| 0838+13 | 3c207 | 80 | Cl | 2.4 | 31 | II | 2 |
| 0840+29 | 4c29.31 | 48 | PB | 4.0 | 22 | | |
| 0843-33 | | 15 | Ha | 1.7 | 20 | | |
| 0850+14 | 3c208 | 63 | PB | 3.5 | 24 | II | 2 |
| 0854+34 | 4c34.30 | 62 | PB | 1.5 | 22 | | |
| 0855+14 | 3c212 | 85 | Cl | 3.0 | 18 | II | 2 |
| 0903+16 | 3c215 | 58 | Cl | 2.0 | 18 | II | 2 |
| 0905+38 | 3c217 | 89 | Cl | 2.8 | 27 | II | 2 |
| 0917+45 | 3c219 | 60 | Cl | 3.5 | 17 | II | 2 |
| 0927+36 | 3c220.2 | 25 | Cl | 2.4 | 27 | | |
| 0931+39 | 4c39.26 | 87 | PB | 1.5 | 22 | | |
| 0936+36 | 3c223 | 85 | Cl | 5.0 | 18 | II | 2 |
| 0938+39 | 3c223.1 | 30 | Cl | 4.0 | 16 | II | 4 |
| 0939+14 | 3c225 | 57 | Cl | 5.0 | 17 | II | 2 |
| 0941+10 | 3c226 | 77 | Cl | 4.5 | 18 | II | 2 |
| 0945+07 | 3c227 | 76 | Cl | 3.1 | 20 | II | 1 |
| 0947+14 | 3c228 | 81 | Cl | 4.0 | 18 | II | 2 |
| 0951+69 | 3c231 | 50 | Mi | 1.7 | 17 | I | 2 |
| 0958+29 | 3c234 | 68 | Cl | 3.0 | 17 | II | 2 |
| 1030+58 | 3c244.1 | 49 | Cl | 2.2 | 17 | II | 2 |
| 1040+12 | 3c245 | 78 | Cl | 3.5 | 27 | II | 2 |
| 1048-09 | 3c246 | 72 | Cl | 2.0 | 20 | | |
| 1056+43 | 3c247 | 57 | Cl | 1.4 | 18 | II | 2 |
| 1059-01 | 3c249 | 81 | Cl | 4.5 | 77 | | |
| 1100+77 | 3c249.1 | 88 | Cl | 3.5 | 31 | II | 2 |
| 1106+25 | 3c250 | 78 | Cl | 5.0 | 27 | II | 2 |
| 1107+37 | 4c37.29 | 88 | PB | 3.5 | 22 | II | 7 |
| 1108+35 | 3c252 | 5 | Cl | 1.8 | 17 | II | 2 |
| 1111+40 | 3c254 | 17 | Mi | 1.3 | 18 | II | 2 |
| 1136-67 | | 76 | Cl | 3.0 | 21 | | |
| 1136-13 | OM-161 | 69 | Cl | 2.9 | 36 | | |
| 1137+66 | 3c263 | 70 | Cl | 2.7 | 17 | II | 2 |
| 1140+22 | 3c263.1 | 34 | Cl | 2.2 | 24 | II | 2 |
| 1142+19 | 3c264 | 71 | Cl | 2.0 | 20 | I | 2 |
| 1142+31 | 3c265 | 64 | Cl | 3.6 | 50 | II | 2 |
| 1143-31 | | 86 | Cl | 2.0 | 21 | | |
| 1147+13 | 3c267 | 73 | Cl | 3.5 | 50 | II | 2 |
| 1157+73 | 3c268.1 | 55 | Cl | 4.3 | 19 | II | 2 |
| 1158+31 | 3c268.2 | 64 | Cl | 2.7 | 17 | II | 11 |
| 1203+64 | 3c268.3 | 83 | Cl | 3.2 | 33 | II | 2 |
| 1206+43 | 3c268.4 | 2 | Cl | 2.2 | 24 | II | 2 |
| 1211-41 | | 6 | PB | 2.2 | 25 | | |
| 1216-10 | | 52 | Cl | 2.2 | 20 | | |
| 1216+06 | 3c270 | 8 | Cl | 3.0 | 20 | I | 17,7 |
| 1218+33 | 3c270.1 | 69 | Cl | 2.3 | 27 | II | 2 |
| 1222+13 | 3c272.1 | 36 | Cl | 1.9 | 17 | I | 7 |
| 1222+21 | 4c21.35 | 43 | Cl | 2.8 | 32 | | |
| 1222+42 | 3c272 | 88 | Cl | 4.0 | 27 | II | 2 |
| 1226+02 | 3c273 | 69 | Cl | 3.0 | 41 | | |
| 1228+12 | 3c274 | 89 | Cl | 2.0 | 17 | I | 2 |
| 1232+21 | 3c274.1 | 81 | Cl | 2.8 | 17 | II | 2 |
| 1233+16 | | 72 | Cl | 4.5 | 20 | | |
| 1241+16 | 3c275.1 | 43 | Cl | 2.4 | 27 | II | 2 |
| 1249+09 | | 82 | Cl | 3.0 | 51 | | |
| 1251+15 | 3c277.2 | 62 | PB | 4.0 | 27 | II | 2 |
| 1251+27 | 3c277.3 | 21 | Cl | 1.5 | 19 | I | 15 |
| 1251-12 | 3c278 | 57 | Cl | 1.3 | 20 | I | 8 |
| 1253+37 | 4c37.35 | 78 | PB | 5.0 | 23 | | |
| 1254+47 | 3c280 | 58 | Cl | 1.2 | 18 | II | 2 |
| 1257+38 | 4c38.34 | 77 | PB | 3.0 | 22 | | |
| 1258+40 | 3c280.1 | 86 | Cl | 3.1 | 24 | II | 2 |
| 1301+38 | 4c38.35 | 68 | PB | 2.8 | 16 | II | 16 |
| 1308+27 | 3c284 | 80 | Cl | 3.0 | 17 | II | 2 |
| 1313+07 | | 7 | Cl | 2.0 | 20 | | |
| 1317-00 | 4c00.50 | 89 | Cl | 2.3 | 21 | | |
| 1318+11 | 4c11.45 | 44 | Cl | 1.3 | 27 | | |
| 1319+42 | 3c285 | 4 | Cl | 2.2 | 28 | II | 2 |
| 1328+30 | 3c286 | 37 | Mi | 2.0 | 38 | | |
| 1330+02 | 3c287.1 | 52 | Cl | 1.8 | 20 | II | 7 |
| 1335-06 | 4c-06.35 | 85 | Ha | 2.6 | 32 | | |
| 1343+50 | 3c289 | 78 | Cl | 3.0 | 27 | II | 2 |
| 1350+31 | 3c293 | 53 | Cl | 1.5 | 19 | I | 2 |
| 1352+165 | 3c293.1 | 68 | Cl | 2.3 | 21 | | |
| 1354-17 | op-190.4 | 6 | Cl | 2.7 | 21 | | |
| 1354+19 | 4c19.44 | 84 | Cl | 3.5 | 35 | II | 12 |
| 1358-11 | | 4 | Cl | 5.0 | 29 | II | |
| 1409+52 | 3c295 | 5 | PB | 2.0 | 31 | II | 2 |
| 1413-36 | | 57 | PB | 3.0 | 26 | | |
| 1414+11 | 3c296 | 1 | Cl | 1.6 | 20 | I | 2 |
| 1420+19 | 3c300 | 36 | Cl | 3.0 | 45 | II | 2 |
| 1422+20 | 4c20.33 | 65 | Cl | 3.4 | 37 | | |
| 1423+24 | 4c24.31 | 75 | Cl | 5.0 | 32 | | |
| 1425-01 | 3c300.1 | 37 | Cl | 2.1 | 21 | | |
| 1441+52 | 3c303 | 61 | Cl | 2.2 | 19 | II | 2 |
| 1449-12 | | 87 | Cl | 2.2 | 20 | | |
| 1458+71 | 3c309.1 | 45 | Ha | 2.0 | 33 | | |
| 1502+26 | 3c310 | 54 | Cl | 3.3 | 13 | I | 2 |
| 1508+08 | 3c313 | 55 | Cl | 2.6 | 17 | II | 7 |
| 1511+26 | 3c315 | 72 | Cl | 1.4 | 18 | I | 2 |
| 1512+37 | 4c37.43 | 79 | Cl | 3.5 | 49 | | |
| 1514+00 | 4c00.56 | 28 | Cl | 3.1 | 20 | II | |
| 1522+54 | 3c319 | 45 | Cl | 3.5 | 14 | II | 2 |
| 1529+24 | 3c321 | 64 | Cl | 3.2 | 44 | II | 2 |
| 1529+35 | 3c320 | 46 | PB | 1.5 | 18 | II | 7 |
| 1545+21 | 3c323.1 | 72 | Cl | 3.4 | 19 | II | 11 |
| 1547+21 | 3c324 | 80 | Cl | 5.0 | 39 | II | 2 |
| 1549+62 | 3c325 | 44 | Cl | 1.4 | 18 | II | 2 |
| 1549+20 | 3c326 | 88 | Cl | 5.0 | 18 | II | 2 |
| 1553+20 | 3c326.1 | 56 | PB | 1.5 | 27 | | |
| 1556-21 | | 15 | Cl | 1.7 | 21 | | |
| 1559+02 | 3c327 | 70 | Cl | 2.9 | 20 | II | 4 |
| 1602-63 | | 8 | Cl | 1.9 | 21 | | |
| 1602-09 | | 90 | Cl | 2.8 | 20 | | |
| 1609+66 | 3c330 | 69 | Cl | 2.7 | 17 | II | 2 |
| 1610-608 | | 22 | PB | 3.4 | 54 | II | |
| 1615+32 | 3c332 | 75 | Cl | 3.0 | 19 | II | 7 |
| 1618+17 | 3c334 | 81 | Cl | 3.0 | 27 | II | 2 |
| 1622+23 | 3c336 | 80 | Cl | 2.2 | 18 | II | 2 |
| 1626+27 | 3c341 | 86 | Cl | 3.7 | 17 | II | 2 |
| 1626+39 | 3c338 | 85 | PB | 3.2 | 82 | I | 2 |
| 1627+23 | 3c340 | 87 | Cl | 4.0 | 27 | II | 2 |
| 1627+44 | 3c337 | 36 | Cl | 1.8 | 17 | II | 2 |
| 1634+26 | 3c342 | 78 | Cl | 3.0 | 22 | | |
| 1637-77 | | 41 | PB | 2.6 | 25 | II | 8 |
| 1641+17 | 3c346 | 79 | Cl | 2.0 | 31 | II | 7 |
| 1641+39 | 3c345 | 76 | Ha | 2.2 | 43 | II | 13 |
| 1648+05 | 3c248,HerA | 78 | Ha | 2.2 | 20 | | |
| 1658+47 | 3c349 | 74 | Cl | 3.0 | 17 | II | 2 |
| 1704+61 | 3c351 | 19 | PB | 2.3 | 18 | II | 2 |
| 1709+46 | 3c352 | 77 | Cl | 2.8 | 27 | II | 2 |
| 1717-00 | 3c353 | 5 | Cl | 2.6 | 20 | I | 7 |
| 1723+51 | 3c356 | 30 | Cl | 3.5 | 50 | II | 2 |
| 1726+31 | 3c357 | 61 | PB | 2.9 | 16 | II | 11 |
| 1730-13 | | 26 | Cl | 1.5 | 41 | | |
| 1733-56 | | 83 | PB | 5.0 | 80 | II | 7 |
| 1737-60 | | 85 | PB | 2.6 | 25 | | |
| 1826+74 | 3c379.1 | 50 | Cl | 2.8 | 17 | II | 11 |
| 1832+47 | 3c381 | 67 | Cl | 3.3 | 19 | II | 2 |
| 1836+17 | 3c386 | 13 | Cl | 1.5 | 18 | I | 2 |
| 1842+45 | 3c388 | 63 | PB | 3.2 | 19 | II | 2 |
| 1845+79 | 3c390.3 | 53 | Cl | 5.0 | 17 | II | 2 |
| 1938-155 | OV-164 | 74 | Da | 1.8 | 80 | | |
| 1939+60 | 3c401 | 38 | PB | 1.9 | 19 | II | 2 |
| 1949+02 | 3c403 | 37 | Cl | 2.6 | 20 | II | 4 |
| 2014-55 | | 47 | Cl | 2.2 | 21 | | |
| 2018+29 | 3c410 | 51 | Cl | 1.5 | 44 | | |
| 2019+09 | 3c411 | 60 | Cl | 2.6 | 31 | | |
| 2040-26 | | 2 | Cl | 3.4 | 20 | II | 7 |
| 2058-28 | | 68 | Cl | 3.0 | 20 | | |
| 2104-25 | OX-208 | 39 | Cl | 3.2 | 26 | I | 7 |
| 2104+76 | 3c427.1 | 89 | Cl | 3.4 | 19 | II | 2 |
| 2106+49 | 3c428 | 83 | Cl | 5.0 | 50 | | |
| 2117+60 | 3c430 | 55 | Cl | 2.8 | 19 | II | 4 |
| 2121+24 | 3c433 | 21 | Cl | 2.0 | 18 | I | 2 |
| 2130-53 | | 53 | Cl | 1.1 | 21 | I | 7 |
| 2135-14 | OX-158 | 77 | Cl | 1.7 | 20 | II | 8 |
| 2141+27 | 3c436 | 3 | Cl | 2.9 | 17 | II | 2 |
| 2145+15 | 3c437 | 69 | Cl | 3.5 | 18 | II | 2 |
| 2153-69 | | 67 | PB | 1.5 | 21 | | |
| 2153+37 | 3c438 | 47 | PB | 2.3 | 19 | I | 7 |
| 2203+29 | 3c441 | 32 | Cl | 2.0 | 18 | II | 2 |
| 2211-17 | 3c444 | 12 | Cl | 2.2 | 20 | II | 7 |
| 2212+13 | 3c442 | 83 | Cl | 3.0 | 18 | | |
| 2221-02 | 3c445 | 67 | Cl | 2.7 | 21 | II | 1 |
| 2229+39 | 3c449 | 73 | Cl | 2.2 | 18 | I | 2 |
| 2239+33 | | 74 | PB | 2.8 | 16 | | |
| 2243+39 | 3c452 | 77 | Cl | 5.0 | 17 | II | 2 |
| 2247+11 | | 24 | Cl | 2.0 | 21 | I | 2 |
| 2251+15 | 3c454.3 | 72 | Ha | 2.8 | 43 | | |
| 2252+12 | 3c455 | 80 | Cl | 3.0 | 38 | II | 2 |
| 2310+05 | 3c458 | 23 | Cl | 2.4 | 20 | | |
| 2314+038 | 3c459 | 85 | Da | 3.3 | 52 | II | 7 |
| 2317-27 | | 85 | PB | 2.7 | 25 | II | 7 |
| 2318+23 | 3c460 | 65 | Cl | 4.5 | 24 | II | 2 |
| 2335+26 | 3c465 | 78 | Cl | 4.0 | 17 | I | 2 |
| 2345+18 | 3c467 | 45 | Cl | 3.0 | 31 | | |
| 2352+79 | 3c469.1 | 87 | Cl | 3.2 | 17 | II | 2 |
| 2354-11 | | 62 | Cl | 2.4 | 21 | | |
| 2356-61 | | 66 | Cl | 3.0 | 54 | II | 3 |
| 2356+27 | 4c27.54 | 78 | PB | 4.0 | 22 | | |
| 2356+43 | 3c470 | 86 | Cl | 5.0 | 45 | II | 2 |

In Col.4: Cl-Clarke et al.(1980); Ha-Haves(1975); PB-Birch(1982); Mi-Mitton(1972); Da-Davis et al.(1983)

LITERATURE for radio-maps and FR classes.